\documentstyle[aps,twocolumn,epsf,epsfig,floats,amssymb]{revtex}
\begin{document}

\twocolumn[\hsize\textwidth\columnwidth\hsize\csname
@twocolumnfalse\endcsname

\title{\bf Evaluating the quality of ground-based microwave radiometer \\
measurements and retrievals using detrended fluctuation and \\ 
spectral analysis
methods}

\author{K. Ivanova, E.E. Clothiaux, and H.N. Shirer }

\address{ Department of Meteorology, Pennsylvania State University, University
Park, PA}
\address{kristy@essc.psu.edu}

\author{T.P. Ackerman }
\address{Pacific Northwest National Laboratory, Richland, WA}
  \author{J.C.
Liljegren}
\address{Environmental Research Division, Argonne National 
Laboratory, Argonne, IL}
\author{M. Ausloos }
\address{GRASP and SUPRAS, B5, Universit\'e de Li\`ege}
\address{4000 Li\`ege, Belgium}

\maketitle

\begin{abstract}

Time series both of microwave radiometer brightness temperature measurements at
23.8 and 31.4 GHz and of retrievals of water vapor and liquid water path from
these brightness temperatures are evaluated using the detrended fluctuation
analysis method. As quantified by the parameter $\alpha$,
this method (i) enables identification of the time scales over which noise
dominates the time series and (ii) characterizes the temporal range of
correlations in the
time series. The more common spectral analysis method is also used to 
assess the
data and its results are compared with those from detrended 
fluctuation analysis
method. The assumption that measurements should have certain scaling properties
allows the quality of the measurements to be characterized. The additional
assumption that the scaling properties of the measurements of an atmospheric
quantity are preserved in a useful retrieval provides a means for 
evaluating the
retrieval itself. Applying these two assumptions to microwave radiometer
measurements and retrievals demonstrates three points.
First, the retrieved water vapor path during cloudy-sky periods can 
be dominated
by noise on shorter than ~30~min time scales ($\alpha$-exponent = 0.1)
and exhibits no
scaling behavior at longer time scales. However, correlations in the brightness
temperatures and liquid water path retrievals are found to be consistent with a
power-law behavior for time scales up to 3 hr with an 
$\alpha$-exponent equal to
approximately 0.3, as in other geophysical phenomena. Second, clear-sky, {\it
moist} atmospheres show the expected scaling for both measurements 
and retrievals
of the water vapor path. Third, during clear-sky, {\it dry} atmospheric
days, instrument noise
from the 31.4 GHz channel compromises the quality of the water vapor path
retrieval. The detrended fluctuation analysis method is thus proposed as means
for
assessing the quality of both the instrument data and the retrieved parameters
obtained from these data.

\end{abstract}

]
\narrowtext

\setcounter{section}{0}

\section{Introduction}

The U.S. Department of Energy (DOE) Atmospheric Radiation Measurement (ARM)
program (Stokes and Schwarts 1994) now operates two-channel microwave 
radiometers
(Liljegren et al. 1994) at its Southern Great Plains (SGP), Tropical Western
Pacific (TWP), and North
Slope of Alaska (NSA) sites. The DOE ARM microwave radiometers operate at
frequencies
of 23.8 and 31.4 GHz and are used to retrieve column water vapor and 
liquid water
amounts (Liljegren and Lesht 1996). The range of atmospheric conditions across
the
three DOE ARM sites is extreme, raising questions concerning the reliability of
the measurements and the retrievals. For these reasons, Liljegren (1999) has
implemented an automated scheme for calibrating the instrument radiances at
the different sites that dramatically improves their quality. 
Moreover, Liljegren
et al. (2000) have developed a new retrieval for water vapor and liquid
water path that attempts to remove biases resulting from changes in atmospheric
conditions from one measurement to the next. Additional approaches to 
retrieving
accurate estimates of column water vapor and liquid water paths currently are
being pursued, such as the Bayesian approach adopted by McFarlane et 
al. (2000).

An important motivation for the efforts of Liljegren et al. (2000) 
and McFarlane
et al. (2000) is that the accuracy of the existing retrieval of cloud liquid
water path (Westwater 1993; Liljegren and Lesht 1996) is not
sufficient for many emerging studies on the interaction between stratus clouds
and
atmospheric radiation. In particular, biases over several-hour periods in the
existing
liquid water path retrieval frequently occurr, rendering their use in 
atmospheric
radiation transfer studies difficult at best. Using a specially 
designed analysis
technique that is insensitive to these biases, we concentrate in this study on
characterizing the quality of the measurements and retrievals based 
on assessment
of the correlations between fluctuations in these signals.
The emphasis here is on analyzing the fluctuations and ascertaining 
the temporal
range
over which they are correlated. This range is limited by the 
discretization step
of the measurements and by the inherent lifetimes of cloudy or clear-sky
atmospheric conditions.

Often the nonlinear processes at work in the atmosphere produce data series of
such complexity that traditional statistical and spectral analyses techniques
fail to extract meaningful physical information. Better techniques are clearly
required. Recently, there have been several studies demonstrating 
that long-range
power-law correlations (Peng et al. 1992) can be analyzed to give relevant time
scales in, generally called, {\it self-organized critical} systems
(Bak et al. 1989).

We study both the statistical properties of microwave radiometer 
measurements, as
given by the 23.8 and 31.4 GHz channel brightness temperatures ${T_B}_1$ and
${T_B}_2$, and the liquid water path (LWP) and water vapor path (WVP) retrieved
from them. As a means of assessing the validity of the retrievals, we determine
how well they preserve the correlations of the fluctuations that 
initially exist
in the measurements. We apply the {\it detrended fluctuation analysis} (DFA)
method (Peng et al. 1994) to characterize the correlations in the
signals. This method has been successfull in the investigation of long-range
memory effects in biological (Stanley et al. 1993; Hausdorff et al. 1995)
and financial (Vandewalle and Ausloos 1997; Ausloos and Ivanova 1999; 
Ausloos et
al. 1999; Ausloos et al. 2000) data fluctuations, as well as to study
long-range power-law correlations in the Southern Oscillation Index 
(Ausloos and
Ivanova 2000) and atmospheric signals to identify stratus cloud breaking
(Ivanova and Ausloos 1999a; Ivanova et al. 2000).

We use the DFA to show that the water vapor path retrieval destroys the
correlations in
the brightness temperature measurements
for a cloudy atmosphere and we propose an explanation for this (numerical)
destruction. In addition, we characterize the contributions of instrument noise
to the measurements during dry and moist clear-sky conditions.

Finally, we contrast DFA method results to those from spectral 
analysis (Davis et
al. 1994; Davis et al. 1996; Ivanova and Ackerman 1999) of the same signals,
illustrating the strength of the DFA method relative to the spectral method for
nonstationary conditions typifying the atmosphere.

The paper is organized as follows. In section 2 we describe the data 
sets and the
retrieved quantities that are analyzed. Section 3 sketches the methods of
analysis used, while the results are presented in section 4. The main 
findings of
the study are discussed in section 5 and summarized in section 6 where we also
suggest directions for future work.

\section{Data Analyzed}

The data used in this study are the radiances, recorded as brightness
temperatures,
measured with a Radiometrics Model WVR-1100 microwave radiometer at frequencies
of
23.8 and 31.4 GHz and the vertical column amounts of cloud liquid 
water and water
vapor retrieved from these measurements (Westwater 1993). The microwave
radiometer is equipped with a Gaussian-lensed microwave antenna whose 
small-angle
receiving cone is steered with a rotating flat mirror (http://www.arm.gov). The
data for this particular study are from the microwave radiometer located at the
DOE ARM SGP central facility. For all four case study periods the microwave
radiometer
operated in vertically pointing mode. In this mode the radiometer makes
sequential
1 s radiance measurements in each of the two channels while pointing vertically
upward into the atmosphere. After collecting these radiances the radiometer
mirror is rotated to view a blackbody reference target. For each of the two
channels the radiometer records the radiance from the reference immediately
followed by a measurement of a combined radiance from the reference and a
calibrated
noise diode. This measurement cycle is repeated once every 20 s.

A shorter measurement cycle does not necessarily lead to a larger number of
independent samples. For example, clouds at 2 km of altitude moving 
at 10 $\rm m
\,s^{-1}$ take 15 s to advect through a radiometer field-of-view of 
approximately
5$^\circ.$ Note that the 1 s sky radiance integration time ensures that the
retrieved quantities correspond to a specific column of cloud
above the instrument, as opposed to some longer time average of the cloud
properties in the column above the instrument.

The field of view of the microwave radiometer is 5.7$^\circ$ at 23.8 GHz and
4.6$^\circ$ at 31.4 GHz. The brightness temperature is measured with a random
error of approximately $\pm 0.5$K.
The atmosphere is not optically thick at these two microwave radiometer
frequencies during cloudy conditions. Hence, these two frequencies 
can be used to
retrieve the total column amounts of water vapor and cloud liquid water. The
error for the liquid water retrieval is estimated to
be less than about 0.005~g/cm$^2$ (Liljegren et al. 2000). The water vapor path
retrievals are considered to
have an absolute accuracy of better than 5\%.

Three cases are considered in the current study: (i) a six-day stratus cloud
event,
January 9 to 14, 1998, which is an exceptionally long lasting cloudy period for
the
Southern Great Plains site, that allows us to analyze an unusually long time
series of 25772 data points, (ii) a cloudless dry day, September 29, 1997,
which has 3812 data points and (iii) a cloudless moist day, September 18, 1997,
which
has 4296 data points.

\section{Data Analysis Techniques}

We use two techniques of analysis: the detrended fluctuation analysis 
($DFA$) and
the more common technique of standard spectral analysis. The $DFA$ technique
(Peng et al. 1994) begins with dividing a random variable sequence $y(t)$ of
length $N$
into $N/\tau$ non-overlapping boxes, each containing $\tau$ points. Then, the
local trend (assumed to be linear in this investigation) $z(t)=at+b$ 
in each box
is computed using a linear least-squares fit to the data points in that
box. It is worth noting that the trend assumption can be generalized 
without any
difficulty (Vandewalle and Ausloos 1998) and without giving rise to drastic
changes in the output results. This local trend removal should be contrasted to
that of the common spectral analysis, in which the trend is sometimes removed
over the entire time series.

We define the mean square difference over an interval of length $\tau$ as

$$ F^2(\tau) = {1 \over \tau} {\sum_{t=k\tau+1}^{(k+1)\tau}
{\left[y(t)-z(t)\right]}^2} , {\hskip 0.5cm} k=0,1,2, \cdots, ({N \over \tau}
-1).\eqno(1)$$

Averaging $F^2(\tau)$ over the $N/\tau$ intervals gives the 
DFA-function $\langle
F^2(\tau) \rangle$ that describes the correlation in fluctuations as a function
of
$\tau$. The procedure is repeated for almost all realistic 
$\tau$-interval sizes
and the behavior of the DFA-function is expected to follow a power law (Peng et
al. 1994)
$$\langle F^2(\tau) \rangle^{1/2} \sim \tau^{\alpha}.\eqno(2)$$ {\noindent An
exponent $\alpha \not= 1/2$ over a certain range of $\tau$ values
implies the existence of long-range correlations in that time 
interval. In fact,
the exponent $\alpha$ is equal to the Hurst exponent of the rescaled
range (R/S) method, often denoted $H_1$, or $H$, in the literature 
(Hurst et al.
1965).
An $\alpha < 1/2$ indicates antipersistence (Addison 1997), meaning that an
increment in a signal is most likely followed by a decrement in the signal and
vice versa. In contrast, $\alpha > 1/2$ indicates persistence whereby 
increments
in a
signal are most likely followed by additional increments, and decrements in a
signal are most likely followed by additional decrements. For $\alpha=0.5$, an
increment in a signal is equally likely to be followed by either an 
increment or
decrement and similarly for a decrement. Random Brownian motion signals are
characterized by an exponent $\alpha=0.5$, while Gaussian white noise sequences
have $\alpha = 0$. For fractional Brownian motion, $\alpha \neq 0.5$ (Turcotte
1997; Addison 1997).

To relate the DFA results to more common analysis techniques, we 
investigate the
high-frequency behavior of the fluctuations by applying the standard spectral
analysis method (Panter 1965). The power spectral density $S(f)$ of an (assumed
self-affine) time series $y(t)$ has a power-law dependence on frequency $f$,
$$S(f) \sim {{1}\over{f^\beta}},\eqno(3)$$

\noindent
following from the Fourier transform of the signal (Panter 1965). If the two
scaling exponents $\alpha$ and $\beta$ are well defined, then the 
Wiener-Khinchin
relation $\beta = 2 \alpha +1$ holds for $0<\alpha<1$ ($1<\beta<3$) (Turcotte
1997; Monin and Yaglom 1975; Heneghan and McDarby 2000). In terms of the
exponents ($\alpha$ and $\beta$) of the signal, we can talk about pink noise
$\alpha=0$
($\beta=1$),
brown noise $\alpha=1/2$ ($\beta=2$) or black noise $\alpha>1/2$ ($\beta>2$)
(Schroeder 1991). Black noise is related to persistence.
In contrast, inertial subrange turbulence for which $\beta=5/3$ gives
$\alpha=1/3$
(Frisch 1995), which places it in the antipersistence regime.

The main advantages of the DFA method over other techniques, such as 
the Fourier
transform (Panter 1965) and the rescaled range (R/S) (Hurst et al. 1965)
methods, are: (i) local and large-scale trends are avoided, and (ii) the upper
scaling limit (here the correlation time) crossover is well
defined. In contrast, detrending of data before applying a Fourier transform
leads to questions about the accuracy of the spectral exponent. For example, we
know that the Fourier transform is inadequate for non-stationary
signals. (An excellent discussion of these aspects may be found in Pelletier
(1997).) Thus, we expect that the DFA method will allow a better understanding
of the noise contribution to the retrieved sequences for different atmospheric
conditions.

Finally, in order to determine how a noise-like sequence alters the scaling
properties
of a signal, we also perform a sensitivity-to-noise study by adding synthetic
Gaussian white noise with a mean of zero and a variance of one to a signal
with well defined scaling properties. Namely, we added Gaussian white noise to
the liquid water path signal for the cloudy-sky case period.

\section{Results}

We first present results from the cloudy-sky case study period to 
demonstrate the
scaling properties expected of LWP retrievals (Ivanova et al. 2000; 
Davis et al.
1994).
A similar analysis of WVP retrieval for the cloudy day yields 
different results.
We next present results from the dry clear-sky case study to illustrate how the
influence of instrument noise is characterized by the analysis technique and to
determine the range of scales over which instrument noise dominates
the statistics of measured and retrieved quantities. The last case we consider
is the moist clear-sky day.

\subsection{Cloudy-sky Atmosphere}

Consider microwave radiometer brightness temperature measurements at 23.8 and
31.4 GHz, and retrieved liquid water path (LWP) and water vapor path 
(WVP) for a
cloudy atmosphere. Data are plotted in Fig. 1.
These measurements were obtained at the ARM Southern Great Plains site during
the period from January 9 to 14, 1998 and consist of $N=25 772$ data points
measured with a time resolution of 20~s. The DFA-functions $\langle F^2(\tau)
\rangle^{1/2}$ for the four time series are shown in Fig. 2. For this case the
short-term fluctuation amplitudes in both the 23.8 and the 31.4 GHz frequency
channels are clearly much
larger than the 0.5 K fluctuations expected from instrument noise (see Fig.
1a,b). Hence, instrumental noise contribution is negligible and the brightness
temperatures ${T_B}_1$ and ${T_B}_2$ for both the 23.8 and the 31.4 
GHz frequency
channels, exhibit well defined scaling properties from 3 to about 150 minutes
with an exponent $\alpha=0.36\pm0.01$ (see Figs. 2a,b and Table 1).

To estimate the correlations of the fluctuations in the liquid water 
retrieval we
plot the DFA-function in Fig. 2d for the liquid water path (LWP) data shown in
Fig. 1d. Part of this result has been presented in one of our previous studies
(Ivanova et al., 2000). We view and discuss it here in the framework 
of both the
direct radiance data ${T_B}_1$ and ${T_B}_2$, and the subsequent retrievals of
liquid water path and water vapor path. Specifics of the retrieval
method are discussed in section 5. Similar to the measured brightness
temperatures
${T_B}_1$ and ${T_B}_2$, the DFA-function for the LWP follows a power 
law with an
exponent $\alpha = 0.36 \pm 0.01$ holding over about two decades in time, i.e.,
from 3 to 150 minutes. The correlation coefficient is $R=0.997$. A crossover to
Brownian-like motion with $\alpha=0.47 \pm 0.03$ is readily seen for longer
correlation times for ${T_B}_1$, ${T_B}_2$ and LWP data in Figs. 2a,b,d.

One should note that the lower limit of the scaling range is determined by the
resolution and discretization step of the measurements. Because a 
cloud moves at
a typical speed of 10~m/s and the radiometer is always directed 
upward toward the
same point in the atmosphere, the 20~s discretization step is chosen to ensure
ergodic sampling over the $5^{\circ}$ observation angle of the instrument.
The upper scaling range limit depends on the life time of the cloud system. For
the data in
Fig. 1, the stratus cloud lasts for 6 days, which is an exceptionally long
lasting cloudy period for the Southern Great Plains site, allowing 
our access to
an unusually long time series of 25772 data points.

These results clearly support the hypothesis that power-law 
correlations exist in
the fluctuations of the integrated cloud liquid water for a wide range of time
scales. These power laws are interpreted as signatures of propagation of some
{\it correlated information} across the cloud system for as long as 
150 minutes.
The results for the liquid water path and the two brightness temperature
measurements are also consistent with the calculation of the first order
$H_q|_{q=1}$-exponent; here $H_1=0.29$ as obtained from a separate multifractal
analysis (Marshak et al. 1997, Ivanova and Ausloos 1999b; Kiely and 
Ivanova 1999,
Tessier et al. 1993, Lovejoy et al. 1996) of remotely measured liquid 
water path
(Ivanova and Ackerman 1999) and of directly measured liquid water content
in an oceanic stratus cloud system (Davis et al. 1994).

In contrast, the scaling properties of the retrieved water vapor path are
compromised. The DFA-function (plotted in Fig. 2c) exhibits noise-like
fluctuations from about 3 to about 30~min with an exponent $\alpha\approx 0.10$
(Table 1).
To understand the reason for this behavior in the water vapor column retrieval
for a cloudy atmosphere we examine in section 5 each step of the retrieval
procedure to determine where the change in the correlations of the fluctuations
occurs (Westwater 1978; Liljegren and Lesht 1996).

We relate the DFA results to those given by a spectral analysis of 
the brightness
temperatures
${T_B}_1$ and ${T_B}_2$ and the two retrievals, whose spectra are 
plotted in Fig.
3. The scaling properties of the LWP and the two brightness temperatures are
characterized by exponents $\beta$ that are consistent, within large 
error bars,
with the DFA-$\alpha$ values according to the relation $\beta = 2 \alpha +1$
(Table 2). The WVP spectrum (Fig. 3c) exhibits a noise-like 
contribution at high
frequencies that corresponds to the short time scales given by the DFA-function
in Fig. 2c. The large variability in the power spectra leads to large
uncertainties
in the estimates of the spectral ($\beta$) exponent. This problem of 
the discrete
Fourier transform is known as spectral variance and has been discussed
comprehensively by many authors, among them Priestly (1981) and Percival and
Walden (1994).

\subsection{Dry Clear-sky Atmosphere}

To test whether or not the correlations of the fluctuations are preserved for
other
atmospheric conditions, we analyze the brightness temperature measurements
${T_B}_1$ and ${T_B}_2$, and the retrieved liquid water path and 
water vapor path
for an extremely dry clear-sky (Fig. 4). The precipitable water estimated by
radiosonde measurements for this dry day, September 29, 1997, is of order of
0.9~cm.
These measurements were obtained at the ARM Southern Great Plains site and
consist of
$N=3812$ data points measured with a time resolution of 20~s.

The DFA-functions $\langle F^2(\tau) \rangle^{1/2}$ for these four data series
are shown in Fig. 5. The DFA-function in Fig. 5b illustrates that the 31.4 GHz
brightness temperature channel is dominated by instrumental noise at short
time-scales as a result of both the very low levels of absolute 
humidity and the
absence of any liquid water. As expected, the influence of instrument
noise on the 31.4 GHz channel is particularly severe. The spectral analysis
result
(Fig. 6) can be interpreted along these lines as well. The drop in the value of
$\alpha$ from the measured to the retrieved quantities (Fig. 5, Table 
1) for the
time range from about 3 to 150~min is a result of adding measurement 
noise in the
retrieval
process (section 5). For this dry clear-sky case the LWP signal analysis
demonstrates noise in the
LWP over all time scales, which is expected owing to the absence of 
cloud liquid
water.
The retrieval of small LWP values in the absence of actual cloud 
liquid water is
an artifact
of the retrieval process. This problem is being addressed with an improved
retrieval
algorithm (Liljegren et al. 2000; McFarlane et al. 2000).

To determine how a noise-like sequence alters the scaling properties 
of a signal
with well-defined scaling properties when it is superimposed on a signal,
we performed an idealized sensitivity-to-noise study. We add a 
synthetic Gaussian
white noise with a mean of zero and a variance of one to the liquid water path
data in Fig. 1d. The original signal is anticorrelated with $\alpha=0.36$. The
noise signal is added in successive steps with an amplitude
$\epsilon$ ranging from 1/16 to 1/2 of the LWP signal amplitude. 
Results for the
DFA-function $<F^2(\tau)>^{1/2}$ for two cases, 1/4 and 1/2 noise-to-signal
ratio, are plotted in Fig. 7a as curves (ii) and (iii). The $\alpha$-values for
the noisy signals are plotted in Fig. 7b as a function of the noise-to-signal
ratio $\epsilon$, which can be viewed as representing the degree of noise
contamination. An exponential decrease of the $\alpha$-values relative
to zero-noise-value-$\alpha=0.36$, with $\epsilon$ is observed for the LWP
signal.

In Fig. 7c the respective power spectra are plotted. Note that spectrum labeled
(iii) in Fig. 7c, which is the spectrum of a complex signal obtained from the
superposition
of Gaussian white noise on the LWP signal, is very similar to the 
spectrum of the
WVP retrieval shown in Fig. 3c. However, the DFA-function labeled (iii) in Fig.
7a of the
same complex signal is similar to the DFA-function in Fig. 2c at small time
scales only. The differences at large time scales are related to the fact that
the
data series analyzed are quite different, one being the LWP signal plus white
noise and the other being the WVP retrieval data (Fig. 1c). While spectral
analysis
shows noise in both cases, the detrended fluctuation analysis method finds
differences between the two time series, which emphasize some of the advantages
of the detrended fluctuation analysis over spectral analysis method 
when studying
correlations and anticorrelations of the
fluctuations of a signal.

\subsection{Moist Clear-sky Atmosphere}

September 18, 1997 is an example of a cloudless day that has a very 
high level of
absolute
humidity. The precipitable water on this day is of the order of 4-5~cm. Data
consist of $N=4296$ data points and are plotted in Fig. 8
for the microwave radiometer brightness temperatures ${T_B}_1$ and 
${T_B}_2$, and
the two retrievals. The measured brightness temperatures in both channels have
values that are substantially greater than the cosmic background
plus instrument noise (Fig. 8a,b).
Detrended fluctuation analysis
of the measurements and retrievals highlight this aspect of the data (Fig. 9).
The
23.8 GHz brightness temperature measurements and the retrieved WVP scale with
exponents $\alpha=0.33 \pm 0.02$ and $\alpha=0.31 \pm 0.02$, over 
approximately a
4 hour interval (Table 1). However, the 31.4 GHz brightness temperature
measurements scale with an exponent of $\alpha=0.23 \pm 0.02,$ exhibiting
a higher degree of anticorrelation in the fluctuations as found when there are
relatively higher noise contributions (Fig. 7b). This result is not unexpected,
as the 31.4 GHz channel is more sensitive to cloud liquid water.

The spectral analysis results are plotted in Fig. 10. Spectral 
scaling exponents
$\beta$ estimated for the frequency range consistent with the lower scale range
of the DFA-function for the two brightness temperature measurements
and the WVP retrieval. They are summarized in Table 2. Each $\beta$ exponent is
found to be in agreement within the error bars, with the corresponding $\alpha$
value for the same data via the theoretical relation $\beta=2\alpha+1$) (Table
1).
The flat, high-frequency parts of the spectra in Fig. 10 for periods 
below 1~min,
illustrate the contribution of noise even to measurements and retrievals under
moist, clear-sky atmospheric conditions. The very high frequency noise for time
periods below 1~min is better identified by the
spectral analysis then by the DFA method.

\section{Discussion}

The loss of scaling in the WVP retrievals for the cloudy-sky case study period
was not anticipated and requires an explanation. We now briefly outline the WVP
and LWP
retrieval scheme (Westwater 1978; Liljegren and Lesht 1996) in order 
to identify
the step
in the retrieval at which noise-like behavior is introduced into the WVP
retrieval.

If the total atmospheric transmission at frequency $\nu$ is written as
$\exp{(-\tau_{\nu})}$, then measurements of microwave radiometer brightness
temperature $T_{{\rm B}\nu}$ can be converted to opacity $\tau_{\nu}$ by
$$\tau_{\nu} = \ln\left[{ {(T_{\rm mr} - T_{\rm c})} \over {(T_{\rm 
mr} - T_{{\rm
B}\nu})} }\right],\eqno(4a)$$
where $T_{\rm c}$ is the cosmic background ``big bang'' brightness temperature
equal to 2.8~K and $T_{\rm mr}$ is an estimated ``mean radiating 
temperature'' of
the atmosphere. (Note that Eq. 4a holds for nonprecipitating clouds, 
i.e., clouds
having drops sufficiently small that scattering is negligible.)
Writing $\tau_{\nu}$ in terms of atmospheric constituents, we have 
$$\tau_{\nu} =
\kappa_{{\rm V}\nu}V + \kappa_{{\rm L}\nu}L + \tau_{{\rm d}\nu},\eqno(4b)$$
where $\kappa_{{\rm V}\nu}$ and $\kappa_{{\rm L}\nu}$ are path-averaged mass
absorption coefficients of water vapor and liquid water and $\tau_{{\rm d}\nu}$
is the
absorption by dry atmosphere constituents (i.e., oxygen). Removing the dry
atmosphere
opacity from $\tau_{\nu},$ we have
$$\tau_{\nu}^\ast = \tau_{\nu} - \tau_{{\rm d}\nu} = \ln\left[{ {(T_{\rm mr} -
T_{\rm c})}
\over {(T_{\rm mr} - T_{{\rm B}\nu})} }\right] - \tau_{{\rm d}\nu},\eqno(5)$$

 From measurements at the 23.8~GHz channel, sensitive primarily to water
vapor, and the 31.4~GHz channel, sensitive primarily to cloud liquid water, we
can write two equations for the opacity at the
two frequencies and solve them for the two unknowns $L$ and $V$: $$L =
l_1\tau^\ast_{\nu_1} + l_2\tau^\ast_{\nu_2} \qquad \qquad (LWP)\eqno(6)$$
and
$$V = v_1\tau^\ast_{\nu_1} + v_2\tau^\ast_{\nu_2}, \qquad \qquad 
(WVP)\eqno(7)$$
where
$$l_1 = - \left(\kappa_{{\rm L}\nu_2}{{\kappa_{{\rm V}\nu_1}} \over 
{\kappa_{{\rm V}\nu_2}}} - \kappa_{{\rm L}\nu_1}\right)^{-1},\eqno(8a)$$ 
$$l_2 = \left(\kappa_{{\rm L}\nu_2} - \kappa_{{\rm L}\nu_1}{{\kappa_{{\rm V}
\nu_2}}
\over {\kappa_{{\rm V}\nu_1}}}\right)^{-1},\eqno(8b)$$
$$v_1 = \left(\kappa_{{\rm V}\nu_1} -
\kappa_{{\rm V}\nu_2}{{\kappa_{{\rm L}\nu_1}}\over {\kappa_{{\rm
L}\nu_2}}}\right)^{-1},\eqno(8c)$$
$$v_2 = - \left(\kappa_{{\rm V}\nu_1}{{\kappa_{{\rm L}\nu_2}}\over 
{\kappa_{{\rm
L}\nu_1}}} -
\kappa_{{\rm V}\nu_2}\right)^{-1}.\eqno(8d)$$

To identify the retrieval step that produces changes in the correlations of the
fluctuations,
we perform $DFA$-analysis on time series of $$(T_{\rm mr} - T_{\rm c})/(T_{\rm
mr} - T_{{\rm B}\nu}) \eqno(9)$$ and Eqs. 5, 6, and 7 for the 
cloud-sky case study period.
To understand whether or not
holding $T_{\rm mr}$ constant in Eqs. 5, 6, and 7 is the source of 
the noise, we
generate two time series from each of Eqs. 5, 6, and 7 (Table 3). In the first
time series for
each equation we hold $T_{\rm mr}$ fixed at its monthly-averaged climatological
value (Table 4),
as in the retrieval, while for each of the second time series we estimate a new
value of
$T_{\rm mr}$ for each 20~s sample using other measurements available at the DOE
ARM SGP site.
We do this for
each of the two frequencies $\nu,$ generating a total of 12 sequences. In these
12 sequences
we kept the absorption coefficients $\kappa_{{\rm L}\nu}$ and $\kappa_{{\rm
V}\nu}$ ($\nu=\nu_1,
\nu_2$) fixed. The new scheme of Liljegren et al. (2000) allows for variable
absorption
coefficients and mean radiating temperatures in the retrieval of WVP 
and LWP. So,
we
implement their new scheme and perform a WVP and LWP retrieval with it for this
cloud-sky case study
period. Overall, we generated 14 time series to which we apply the
$DFA$-analysis.

For all 14 data sequences, except the three water vapor signals, the
$DFA$-functions scale with exponent $\alpha=0.36\pm0.01,$ as did the original
measurements of both $T_{\rm B1}$ and $T_{\rm B2}.$ In contrast, each of the
three water vapor path data sequences has the same noise-like, short-range
scaling
behavior. Therefore, the change in short-term fluctuations of retrieved water
vapor path does not result from fixed coefficients and mean radiating
temperatures,
but rather from the transformation of the signals from opacity (Eq. 5) to WVP
(Eq. 7).
Although the transformation from brightness temperature to opacity (Eq. 5)
involves
a non-linear operation, we find that the logarithmic operation preserves the
character of the fluctuations.

To understand why Eqs. 6 and 7, which are similar, exhibit different scaling
properties,
we analyze the relative magnitudes and signs of $l_1$, $l_2$, $v_1$ 
and $v_2$ in
conjunction with the range and fluctuations in the brightness temperature
measurements
$T_{\rm B1}$ and $T_{\rm B2}.$ Atmospheric brightness temperatures 
and opacities,
as
well as their fluctuations, at the frequencies of the two channels 
depend on the
emission
of water vapor, liquid water (for a cloudy atmosphere) and molecular 
oxygen (Fig.
11).
As Fig. 11 illustrates, the 31.4~GHz channel is more sensitive than 
the 23.8~GHz
channel
to variations in liquid water path. In fact, for the cloudy sky the 31.4~GHz
brightness
temperatures $T_{\rm B2}$
fluctuate with amplitudes about twice as large as the amplitudes of 
the 23.8~GHz
brightness
temperatures $T_{\rm B1}$:
$$\delta {T_{\rm B1}} \approx {{1}\over {2}}\delta {T_{\rm 
B2}},\eqno(10a)$$ and
from Eqs. 4 and 5 we have
$$\delta \tau^\ast_{\nu1}\approx {{1}\over {2}}\delta
\tau^\ast_{\nu2}.\eqno(10b)$$
Inspection of the WVP retrieval coefficients in Table 4 shows that 
$$v_1 \approx
-2 v_2.\eqno(10c)$$
Inspecting Eq. 10 in the context of Eq. 7, we see that the contributions of the
fluctuations in
the two channels to WVP are approximately equal in magnitude but 
opposite in sign
and
lead to a numerical phenomenon known as catastrophic cancelation (Blum 1972).
Therefore, the small-scale fluctuations in the WVP sequences during cloudy-sky
period appear as
noise in the $DFA$-function and at high-frequencies in the power spectrum.

During the cloudy-sky period, a comparable problem does not occur in the LWP
retrieval. For the LWP retrieval
$$l_1 \approx -l_2 / 3.5.\eqno(10d)$$
and the 31.4 GHz channel term in Eq. (6) is dominant, preventing catastrophic
cancellation as in Eq. (7) for the cloudy-sky period.

The relative magnitudes
between the two terms in Eqs. (6) and (7) and the influence of instrument noise
on them also explains the dry and moist clear-sky results. For a 
cloud-free sky,
as the column moisture increases, the ratio of the 23.8 GHz to 31.4 GHz
brightness
temperature increases (e.g., compare Figs. 4 and 8), just the opposite to what
happens
on a cloudy-sky day with increasing column liquid (Fig. 11). Consequently, for
clear-sky periods the second term of Eq. (7) becomes increasingly important.

For clear-sky dry periods (section 4b), the small scale fluctuations 
of the 31.4
GHz signal are dominated by instrument noise, because the amplitude levels are
below the thershold for sensitivity to vapor at that frequency, as 
seen from the
microwave radiometer spectrum (Fig. 11). Therefore, the WVP retrieval
algorithm expressed by Eq. (7) becomes noisier.

For a clear-sky moist atmosphere (section 4c) as the ratio of the 23.8 GHz to
31.4 GHz brightness temperature is markedly different than one, Eq. 
10c does not
have such an impact on Eq. (7), the WVP retrieval algorithm, as for cloudy-sky
periods.

The LWP retrieval during clear-sky
periods is always dominated by noise since the second term in Eq. (6) is always
significant.

\section{Summary}

Time series of both microwave radiometer brightness temperature measurements at
23.8 and 31.4 GHz and retrievals of water vapor and liquid water path from the
brightness temperatures are evaluated using both the detrended fluctuation
analysis method and the spectral analysis method. We show that the detrended
fluctuation analysis method is useful for analyzing the power law 
correlations in
such atmospheric signals. In particular, the present analysis
demonstrated the validity of the liquid water path retrieval for a 
cloudy-sky day
and the water vapor path retrieval for a clear-sky moist day.

We found the water vapor path signal to be dominated by noise for a cloudy
atmosphere on time scales less than 30~min. The appearance of noise-like
correlations arises from the specific relationships between the coefficients in
the water vapor retrieval equation, the magnitudes of both the measurements and
fluctuations in the brightness temperature measurements. The latter leads to
catastrophic cancellation of two terms of a similar amplitude and opposite sign
and is related to the sensitivity of the measurements to liquid water 
content at
these frequencies.

On both clear-sky days instrument noise is readily apparent in the 31.4~GHz
channel measurements owing to the absence of emission from liquid
water. In the water vapor path retrieval noise in the 31.4~GHz channel
measurements blurs correlations present in the vapor-sensitive channel
measurements.
For a clear-sky dry atmosphere, instrument noise is readily apparent in the
31.4~GHz channel measurements owing to the absence of liquid water. This noise
artificially blurs the correlations existing in the vapor-sensitive 
channel when
the WVP retrieval is performed.

Since November 1996, the DOE ARM program has been collecting 
coincident microwave
radiometer and millimeter-wave cloud radar data at its SGP site.
We will use detrended fluctuation analysis and spectral analysis methods,
together with other approaches from statistical physics,
to analyze the temporal properties
of boundary layer clouds that have occurred since November 1996. The 
aim of this
next step will be to classify boundary layer clouds according to
their statistical properties and relate these properties to the underlying
dynamics
of the cloud fields. This will lead to better understanding of the impact of
different cloud types on the radiation field.

{\bf Acknowledgments}

This research was supported in part by the Department of Energy through grant
Battelle 327421-A-N4, the Department of Energy through grant number
DE-FG02-90ER61071 and in part by the U. S. Department of Energy, Office of
Science, Office of Biological and Environmental Research, 
Environmental Sciences
Division, under contract W-31-109-Eng-38, as part of the Atmospheric Radiation
Measurement Program. We thank Ed Westwater and Roger Marchand for stimulating
discussions.

\ifdraft\vfill\eject\fi

%\onecolumn

\hsize\textwidth\columnwidth\hsize\csname
@twocolumnfalse\endcsname
\begin{table}[ht] \caption{Values for the DFA $\alpha$-exponent (Eq. 2)
for the microwave radiometer measurements, 23.8 GHz and 31.4 GHz
channel brightness temperatures, and the retrieved water vapor path (WVP) and
liquid water path (LWP)
 for the three cases considered in the text. Here $t_x$
denotes the upper scaling limit for which the $\alpha$ values are obtained.
These $\alpha$ values can be compared to the $\beta$ ones in Table 2 through
the relationship $\beta=2\alpha+1$.} 
\begin{center} \begin{tabular}{||l||c|c|c|c|c||} \hline
$\alpha =$ & $t_x$ [min]& 23.8 GHz & 31.4 GHz & WVP & LWP \\ \hline 
Jan 9-14, 1998
& 150 & 0.36 $\pm$ 0.01 & 0.36 $\pm$ 0.01 & (0.10) [$t_x$=30 min] & 0.36 $\pm$
0.01 \\ cloudy &&&&&\\ \hline Sept 29, 1997 & 150 & 0.21 $\pm$ 0.01 & 
0.10 $\pm$
0.007 & 0.19 $\pm$ 0.01 & 0.06 $\pm$ 0.007 \\ cloudless dry &&&&&\\ \hline Sept
18, 1997 & 240 & 0.33 $\pm$ 0.02 & 0.23 $\pm$ 0.02 & 0.31 $\pm$ 0.02 
& 0.07 $\pm$
0.009 \\ cloudless moist &&&&&\\ \hline \end{tabular} \end{center} \end{table}

\begin{table}[ht] \caption{Values of the spectral exponent $\beta$ (Eq. 3)
for the microwave radiometer measurements, 23.8 GHz and 31.4 GHz
channel brightness temperatures, and the retrieved water vapor path (WVP)
and liquid water path (LWP)
 for the three cases considered in the text. Values of 
$\beta$-exponents are obtained for (i) within time interval [3,1718]~min (Fig. 3);
for (ii) within [2,100]~min (Fig. 6) 
and for (iii) within time interval [2,110]~min (Fig. 10). The
$\beta$-values for the WVP in the (i) and LWP in the (ii) and (iii) cases 
cannot be properly
estimated.} \begin{center} \begin{tabular}{||l|r|c|c|c|c||} 
\hline 
$\beta = $& $N_{data}$ &23.8 GHz & 31.4 GHz & WVP & LWP \\ \hline Jan 9-14, 
1998 &25772&
1.72 $\pm$ 0.04 & 1.71 $\pm$ 0.04 &  & 1.72 $\pm$ 0.04\\ (i) cloudy
&&&&&\\ \hline Sept 29, 1997 &3812 & 1.48 $\pm$ 0.06&0.87 $\pm$ 0.06 
&1.48 $\pm$
0.06& \\ (ii) cloudless dry &&&&&\\ \hline Sept 18, 1997 &4296 &1.70 $\pm$ 
0.07 & 1.55
$\pm$ 0.05 &1.66 $\pm$ 0.06 &\\ (iii) cloudless moist &&&&&\\ \hline \end{tabular}
\end{center} \end{table}

\begin{table}[ht] \caption{Data sets used to trace the changes in the 
correlations analysis}
\begin{center} \begin{tabular}{||r|l||}
\hline
1& 23-GHz Ratio using Monthly-Averaged Mean Radiating Temperature\\
2& 31-GHz Ratio with Monthly-Averaged Mean Radiating Temperature\\
3& 23-GHz Optical Path using Monthly-Averaged Mean Radiating Temperature\\
4& 31-GHz Optical Path using Monthly-Averaged Mean Radiating Temperature\\
5& Water Vapor Path using Monthly-Averaged Mean Radiating Temperature\\
6& Liquid Water Path using Monthly-Averaged Mean Radiating Temperature\\
7& 23-GHz Ratio using Instantaneous Estimated Mean Radiating Temperature\\
8& 31-GHz Ratio with Instantaneous Estimated Mean Radiating Temperature\\
9& 23-GHz Optical Path using Instantaneous Estimated Mean Radiating Temperature\\
10& 31-GHz Optical Path using Instantaneous Estimated Mean Radiating Temperature\\
11& Water Vapor Path using Instantaneous Estimated Mean Radiating Temperature\\
12& Liquid Water Path using Instantaneous Estimated Mean Radiating Temperature\\
13& Water Vapor Path (Variable Coefficients)\\
 14& Liquid Water Path (Variable Coefficients)\\
\hline
\end{tabular} \end{center}
\end{table}

\begin{table}[ht] \caption{Monthly climatological values of the retrieval
coefficients $l_1, \,l_2$ and $v_1, \,v_2$ in Eqs. 6 and 7. Note 
that $r_l=l_2/l_1\approx -3.5$ and $r_v=v_1/v_2\approx -2$}
\begin{center} \begin{tabular}{||c||c|c|c|c||} \hline Month & $l_1$
&$l_2$&$v_1$&$v_2$\\ \hline Jan&-0.144389&0.567176&21.6227&-12.6037\\
Feb&-0.153517&0.554340&21.1463&-12.4573\\
Mar&-0.188830&0.653695&22.1970&-12.9066\\
Apr&-0.258638&0.776852&22.2173&-12.6780\\
May&-0.308970&0.878312&22.7005&-12.8789\\
Jun&-0.290936&0.823459&22.4033&-11.5986\\
Jul&-0.247642&0.751842&22.8299&-13.2155\\
Aug&-0.268197&0.803732&23.1813&-13.4728\\
Sep&-0.261245&0.774673&22.7105&-13.2431\\
Oct&-0.234130&0.731487&22.8530&-13.2600\\
Nov&-0.224780&0.722879&22.0715&-12.7427\\
Dec&-0.143773&0.623870&21.7302&-12.5548\\ \hline \end{tabular} \end{center}
\end{table}

\begin{figure}[ht]
\begin{center}
\leavevmode
\epsfysize=6.5cm
\epsffile{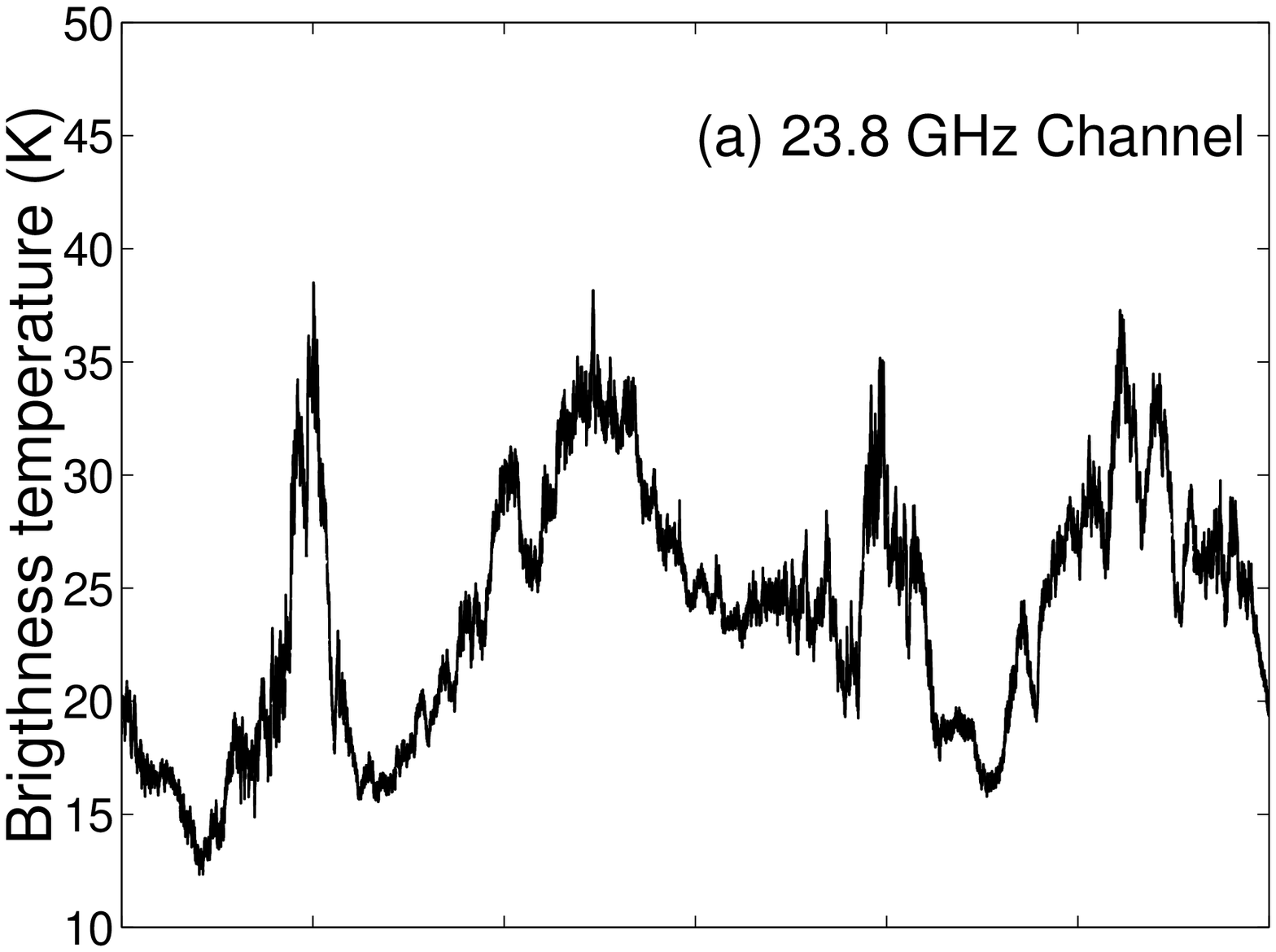}
\hfill
\leavevmode
\epsfysize=6.5cm
\epsffile{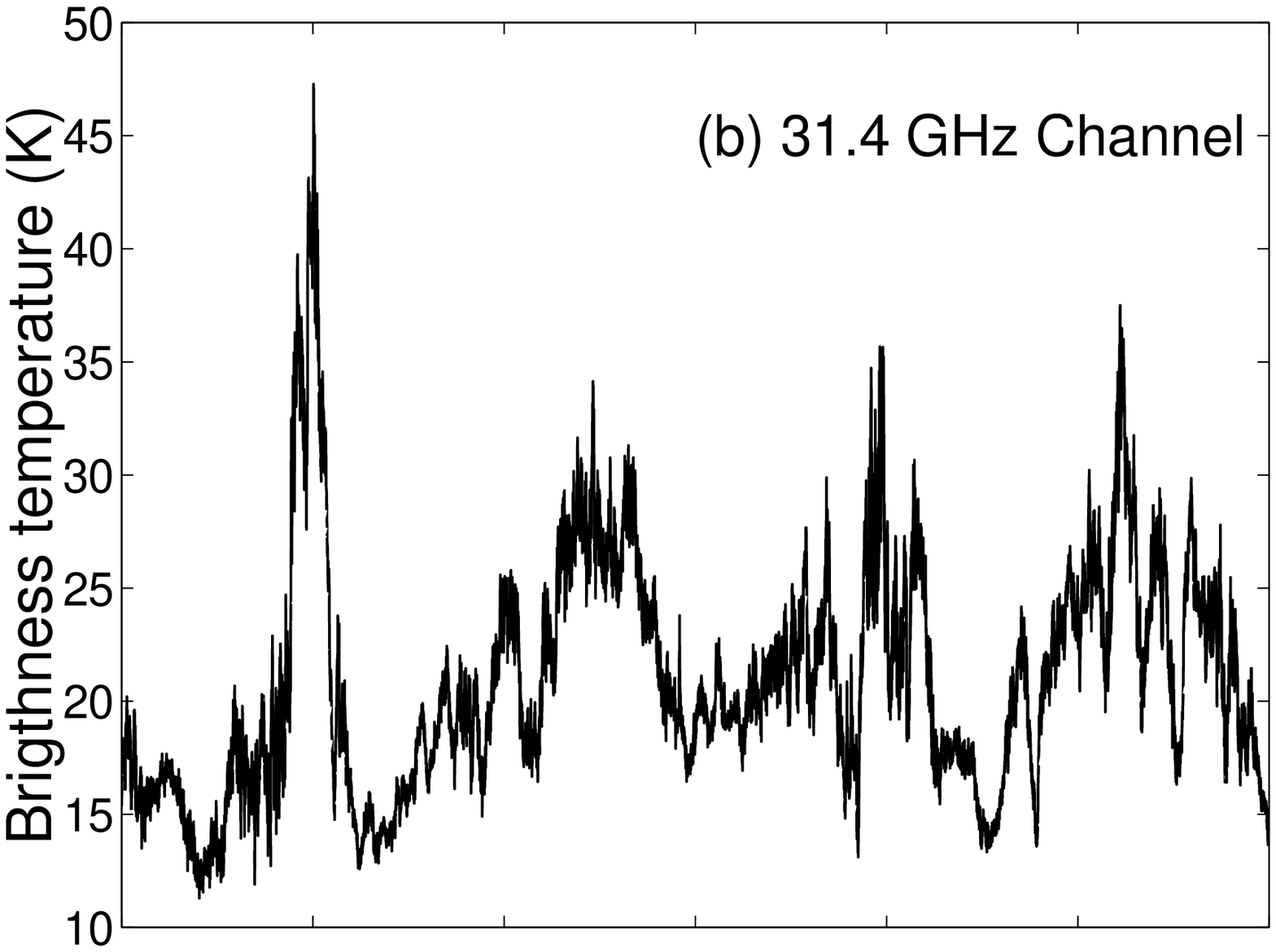}
\vfill
\leavevmode
\epsfysize=6.8cm
\epsffile{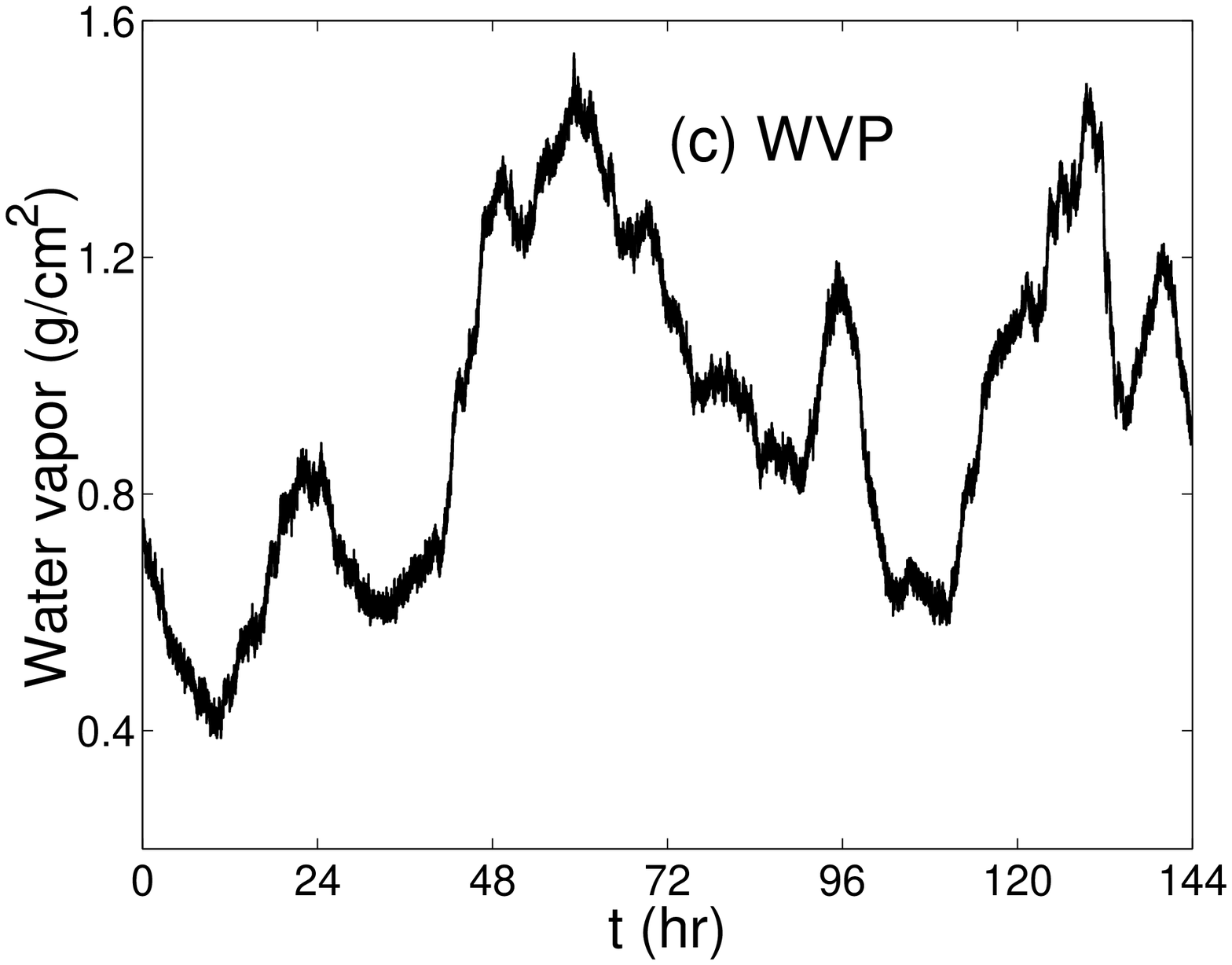}
\hfill
\leavevmode
\epsfysize=6.8cm
\epsffile{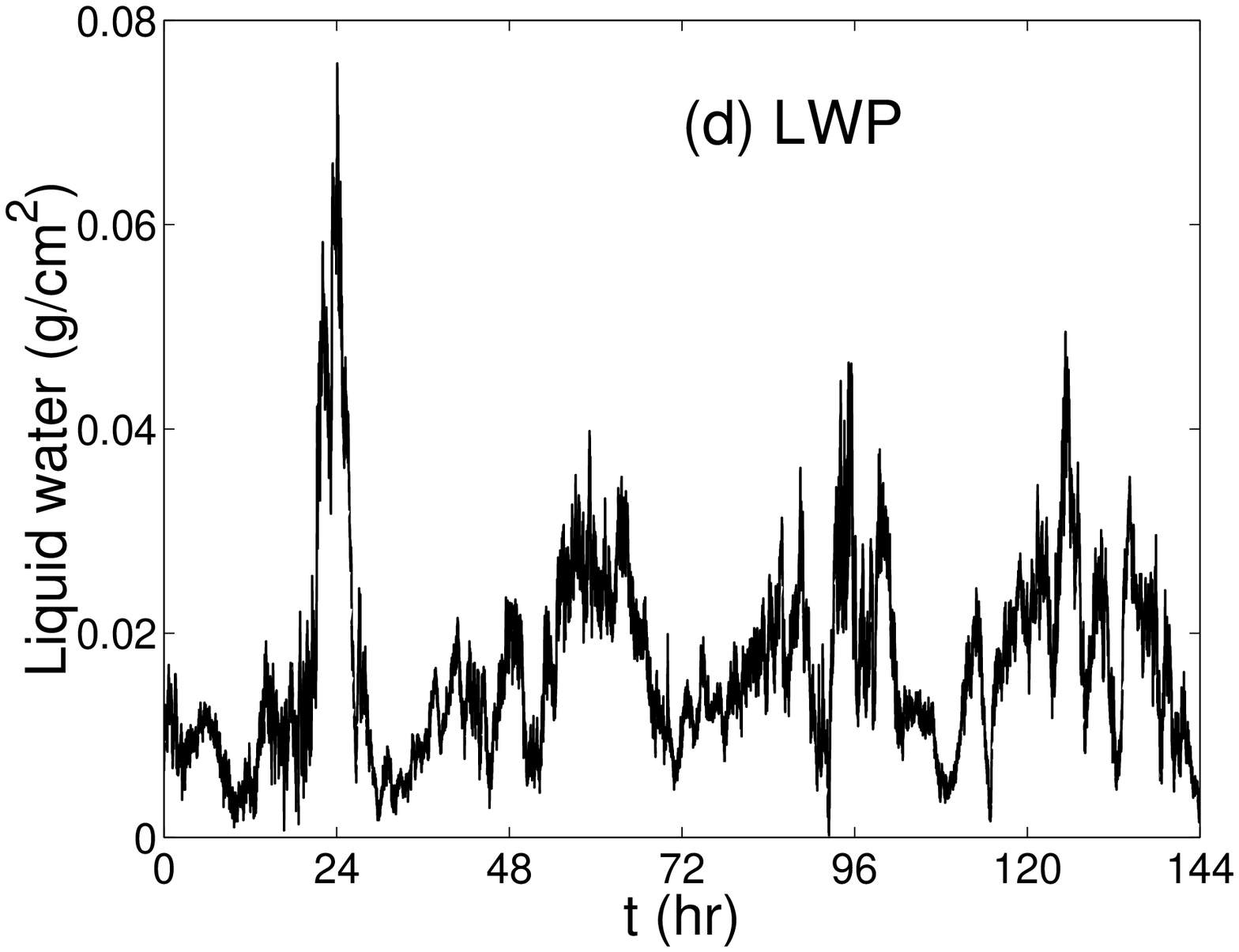}
\end{center}
\caption{Time dependence of the microwave radiometer (a) 23.8 GHz
brightness temperature ${T_B}_1$, (b) 31.4 GHz
brightness temperature ${T_B}_2$, (c) retrieved water vapor path (WVP) 
and (d) retrieved liquid water path (LWP)
for a cloudy case. These measurements 
were obtained
at the ARM Southern Great Plains site with time resolution of 20~s  
during the period from January 9 to 14, 1998. Each time series contains 
$N=25772$ data points.}
\label{fig1}
\end{figure}

\begin{figure}[ht]
\begin{center}
\leavevmode
\epsfysize=6.8cm
\epsffile{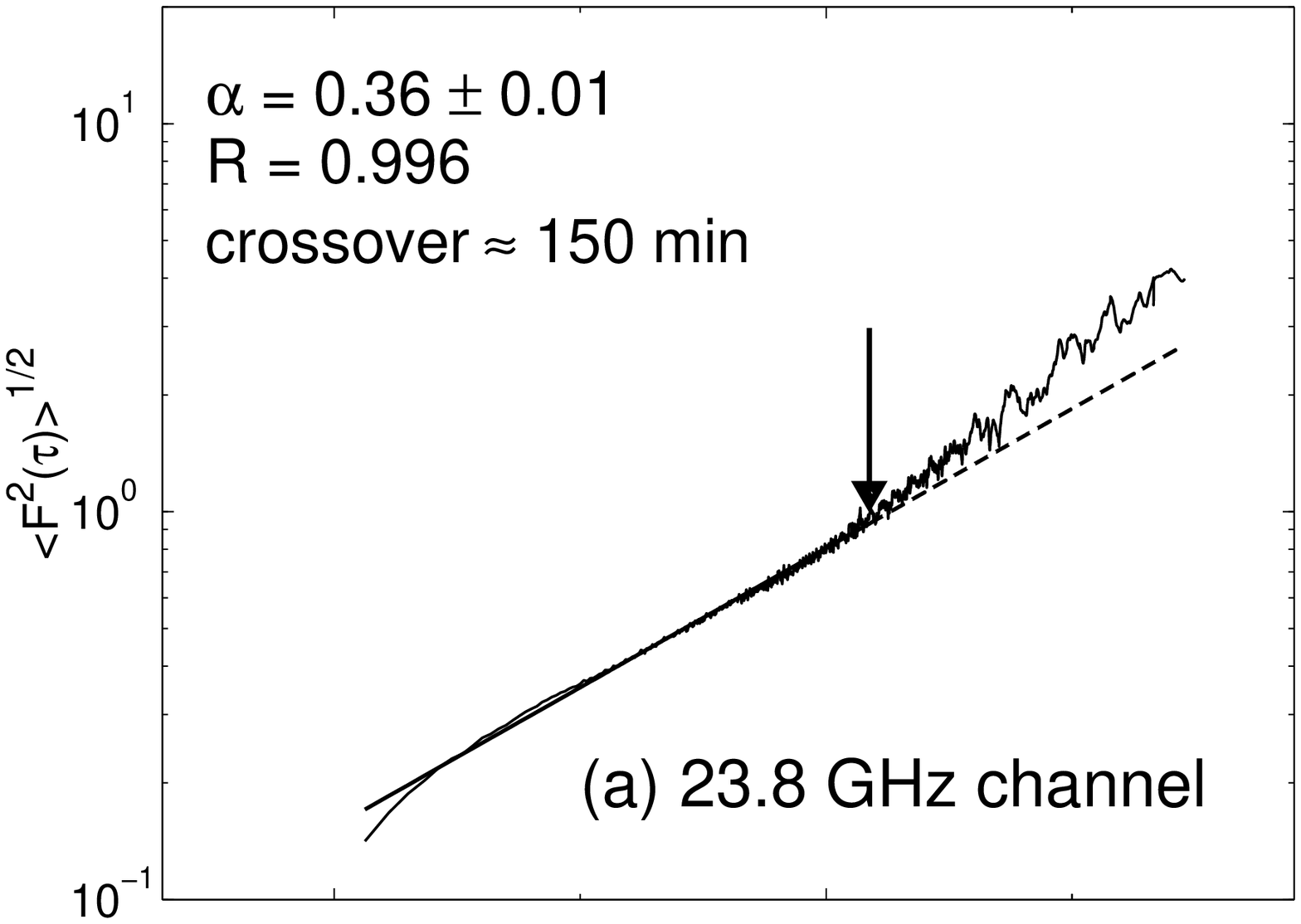}
\hfill
\leavevmode
\epsfysize=6.8cm
\epsffile{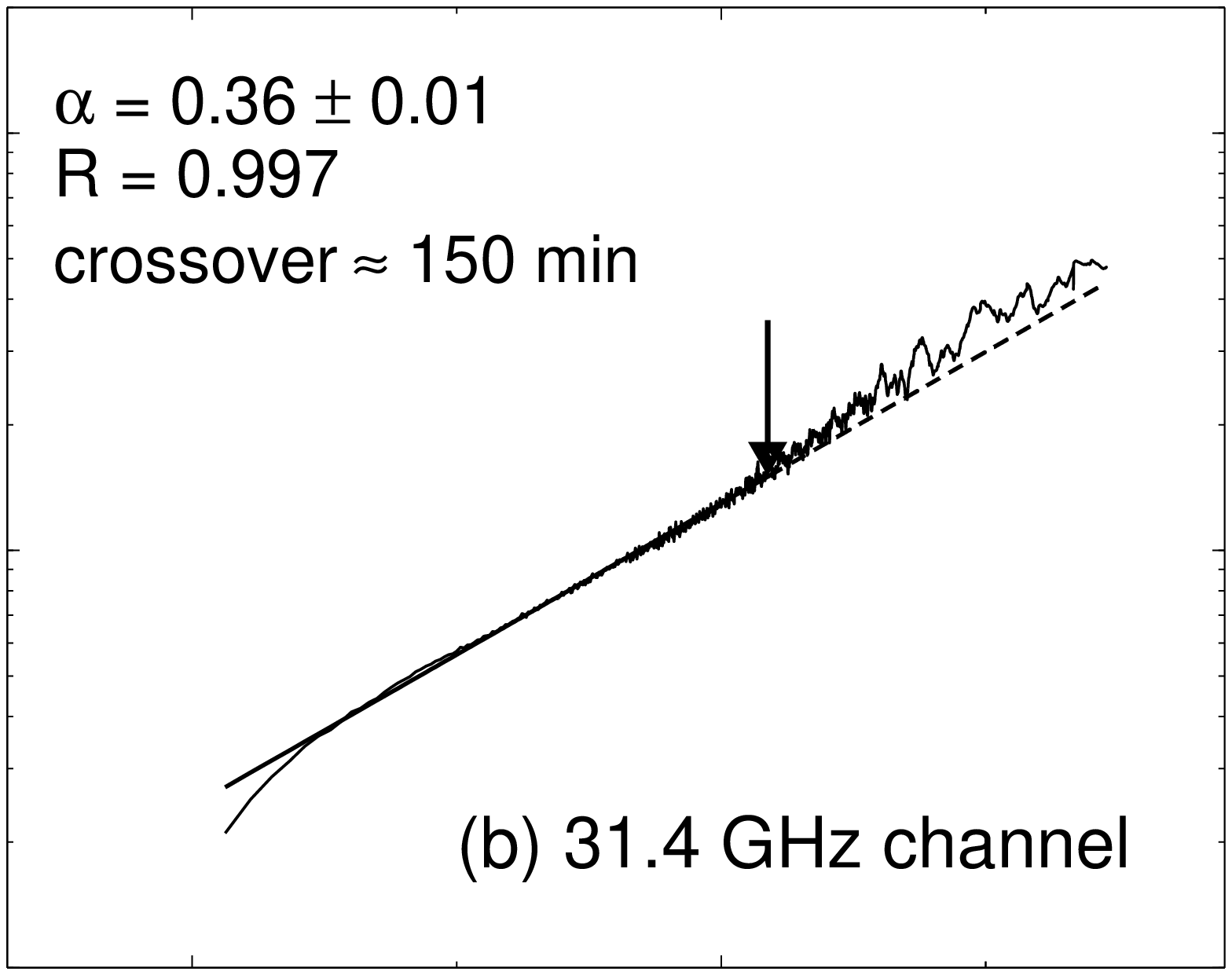}
\vfill
\leavevmode
\epsfysize=7.5cm
\epsffile{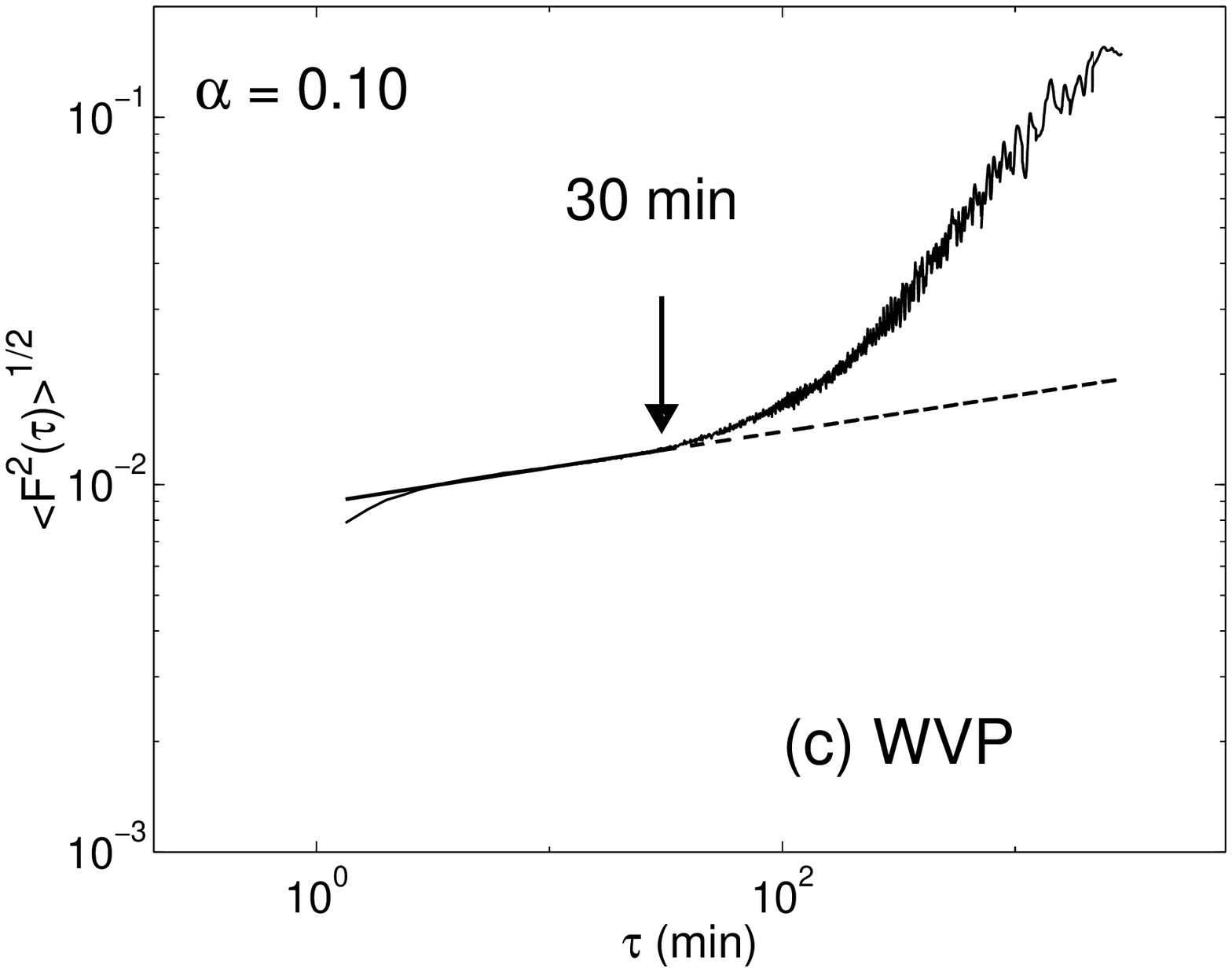}
\hfill
\leavevmode
\epsfysize=7.5cm
\epsffile{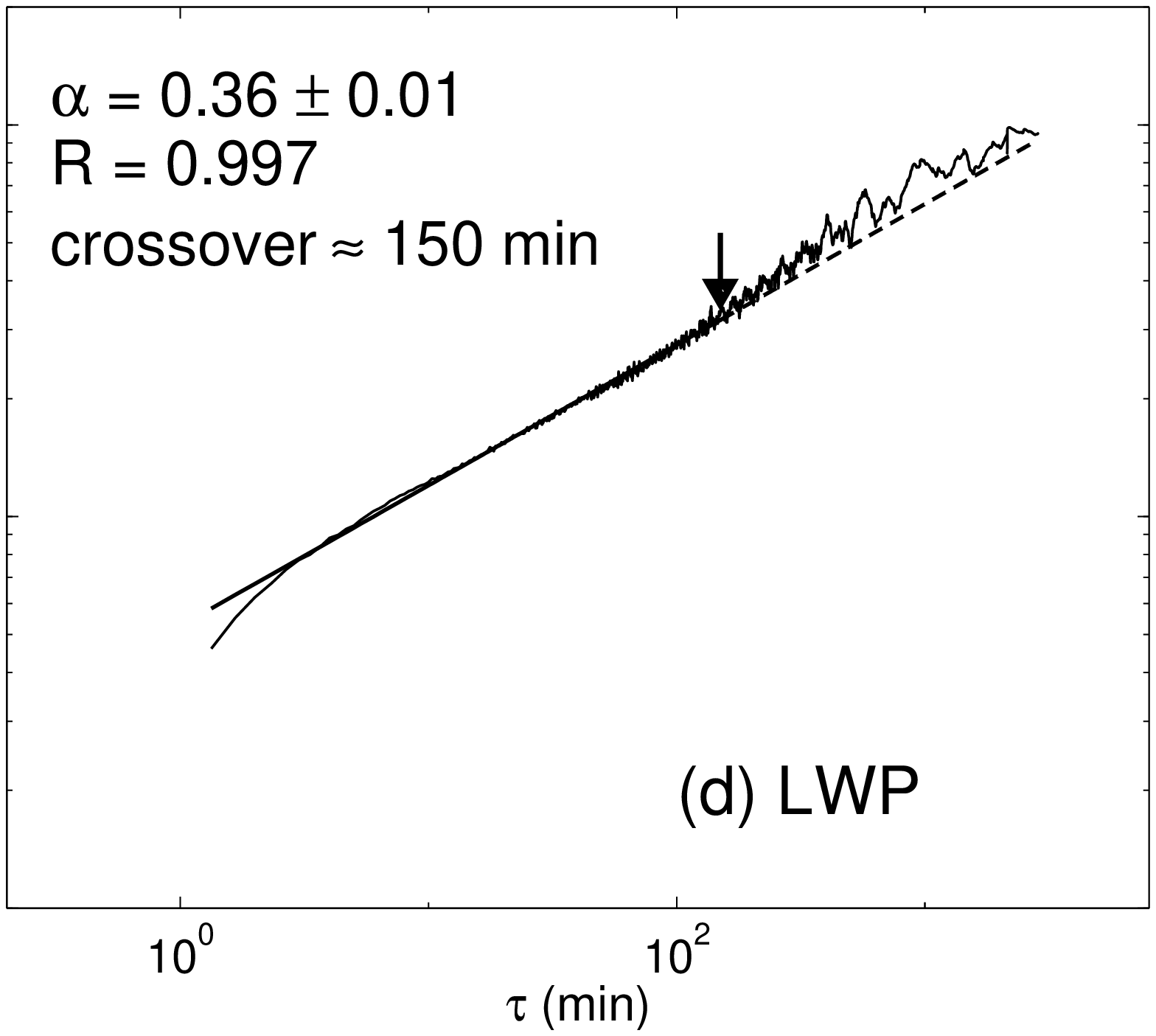}
\end{center}
\caption{The detrended fluctuation analysis (DFA) function 
$<F^2(\tau)>^{1/2}$ (Eq. (1)) for the data in Fig. 1.
Well-defined scaling properties exist in cases (a), (b) and (d) for 
time ranges from about 5 min to about 150 min. However,
the water vapor signal (WVP) in (c) exhibits scaling for a very limited range; 
$\alpha=0.1$ is obtained for the time range from about 3 to about 30~min.}
\label{fig2}
\end{figure}

\begin{figure}[ht]
\begin{center}
\leavevmode
\epsfysize=12cm
\epsffile{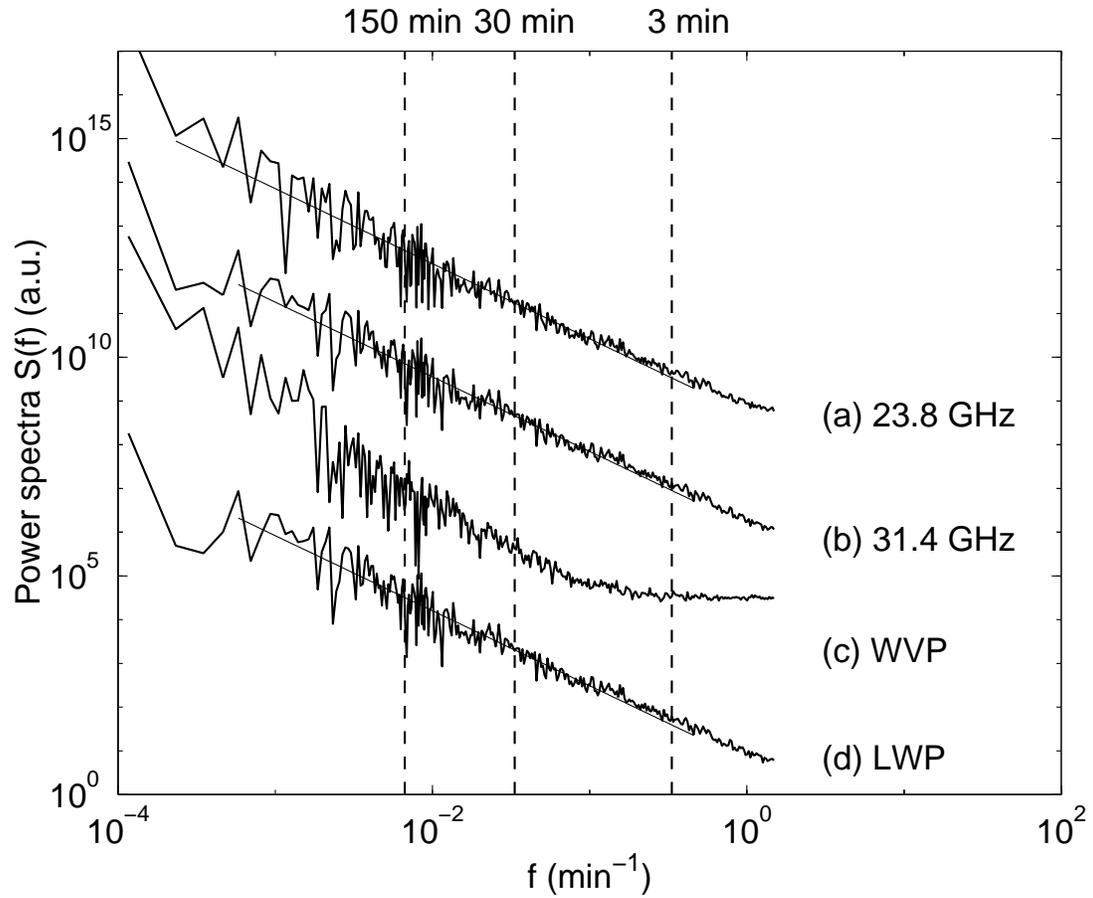}
\end{center}
\caption{Power spectra $S(f)$ for the data in Fig. 1.
Note the consistency in (a), (b) and (d), 
but the WVP spectrum (c) becomes noisy at high frequencies. 
For clarity, the  spectra are offset along the ordinate.}
\label{fig3}
\end{figure}

\begin{figure}[ht]
\begin{center}
\leavevmode
\epsfysize=6.5cm
\epsffile{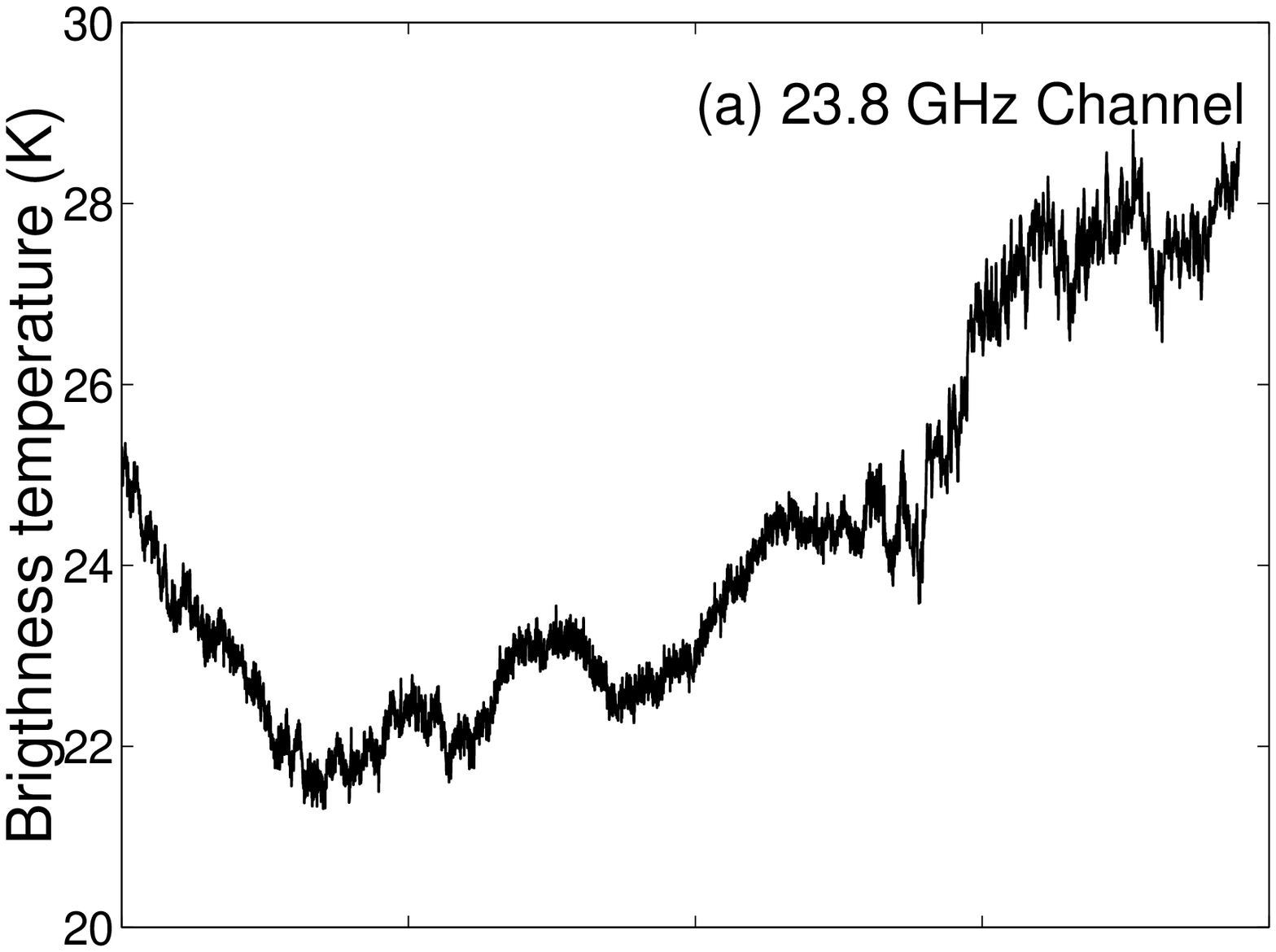}
\hfill
\leavevmode
\epsfysize=6.5cm
\epsffile{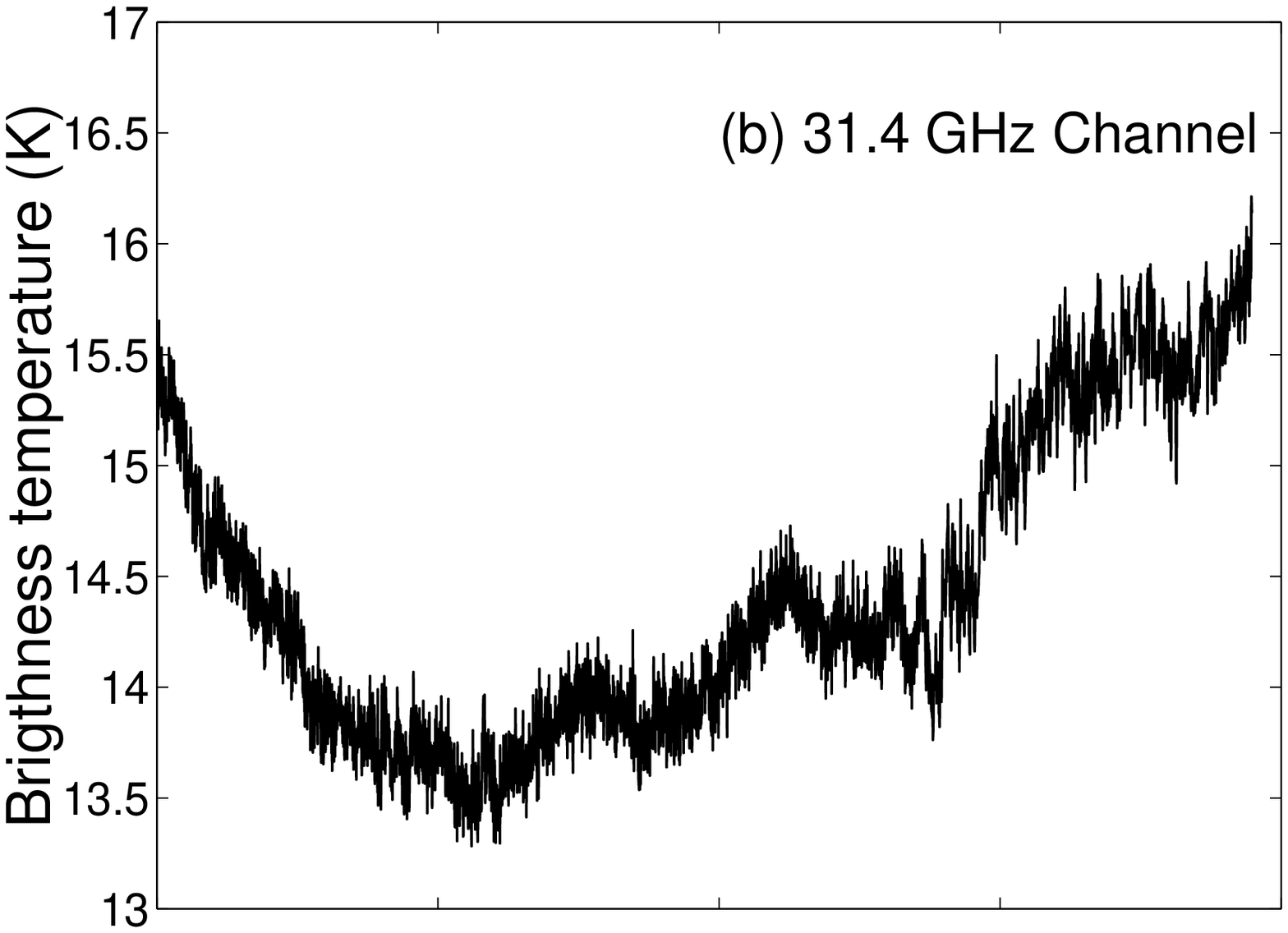}
\vfill
\leavevmode
\epsfysize=7cm
\epsffile{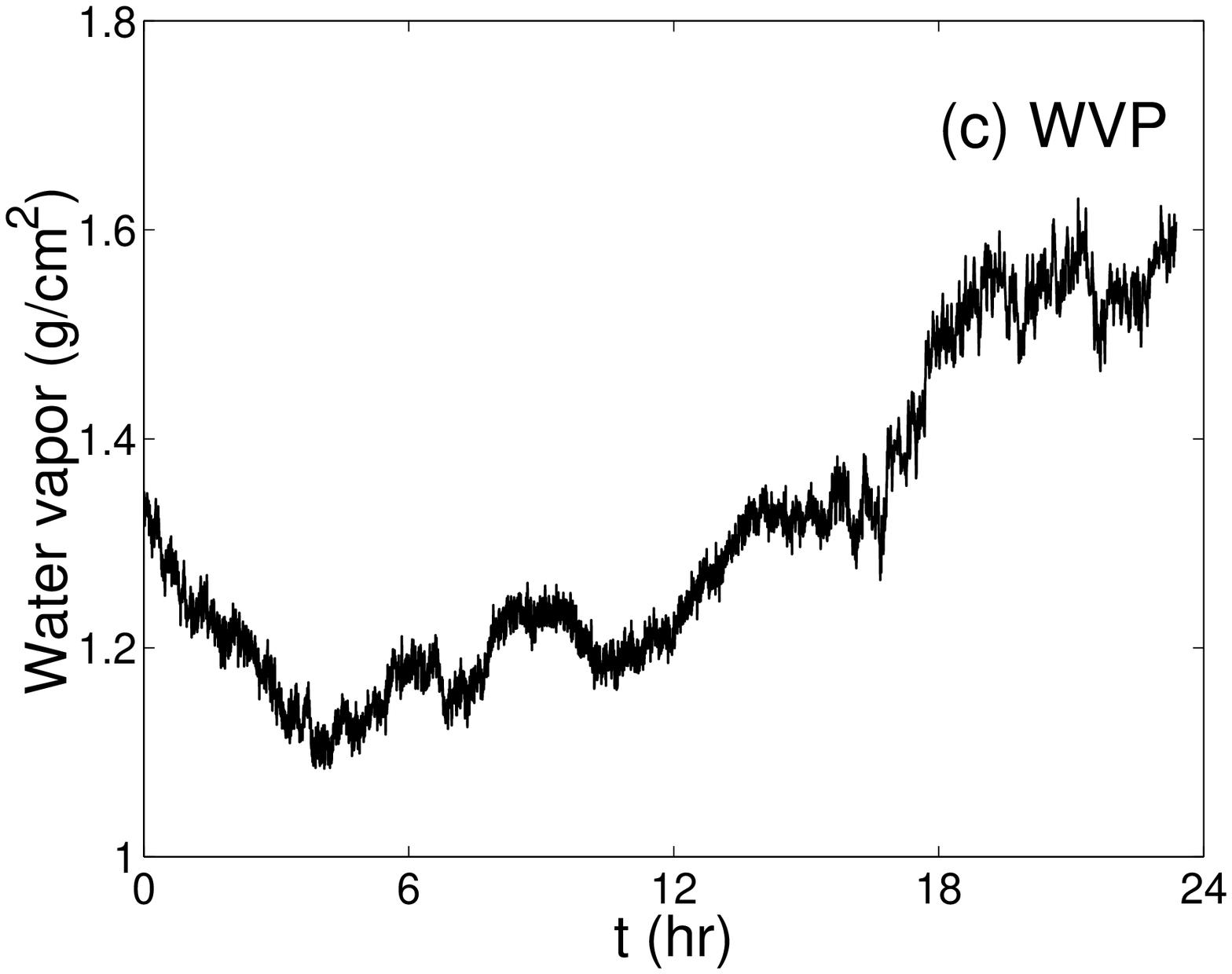}
\hfill
\leavevmode
\epsfysize=7.3cm
\epsffile{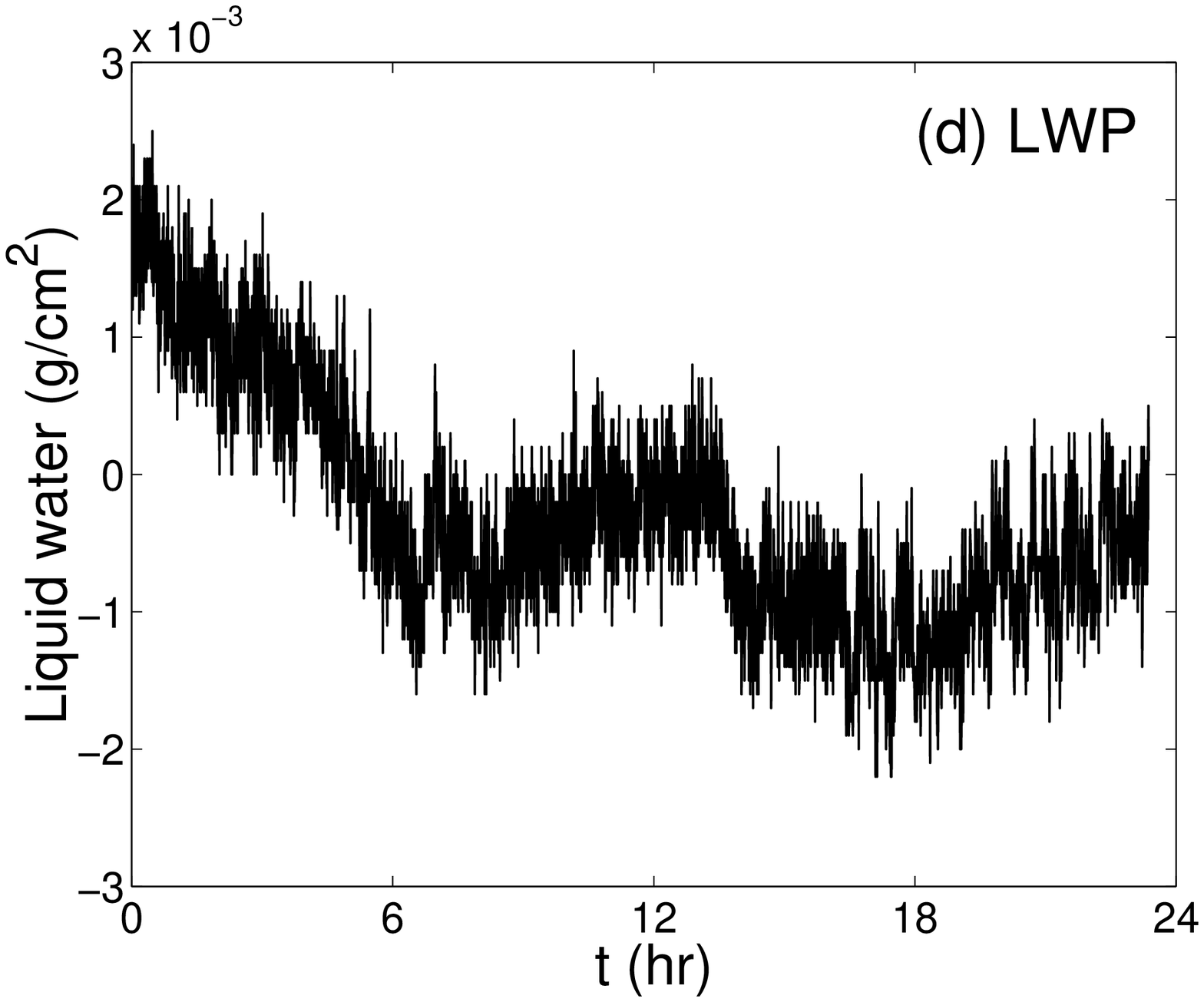}
\end{center}
\caption{Time dependence of the microwave radiometer (a) 23.8 GHz
brightness temperature ${T_B}_1$, (b) 31.4 GHz
brightness temperature ${T_B}_2$, (c) retrieved water vapor path (WVP) 
and (d) retrieved liquid water path (LWP) for a dry cloudless case. 
These measurements were obtained
at the ARM Southern Great Plains site on September 29, 1997.
Each time series contains $N=3812$ data points with a time resolution 
of 20~s.}
\label{fig4}
\end{figure}

\begin{figure}[ht]
\begin{center}
\leavevmode
\epsfysize=12cm
\epsffile{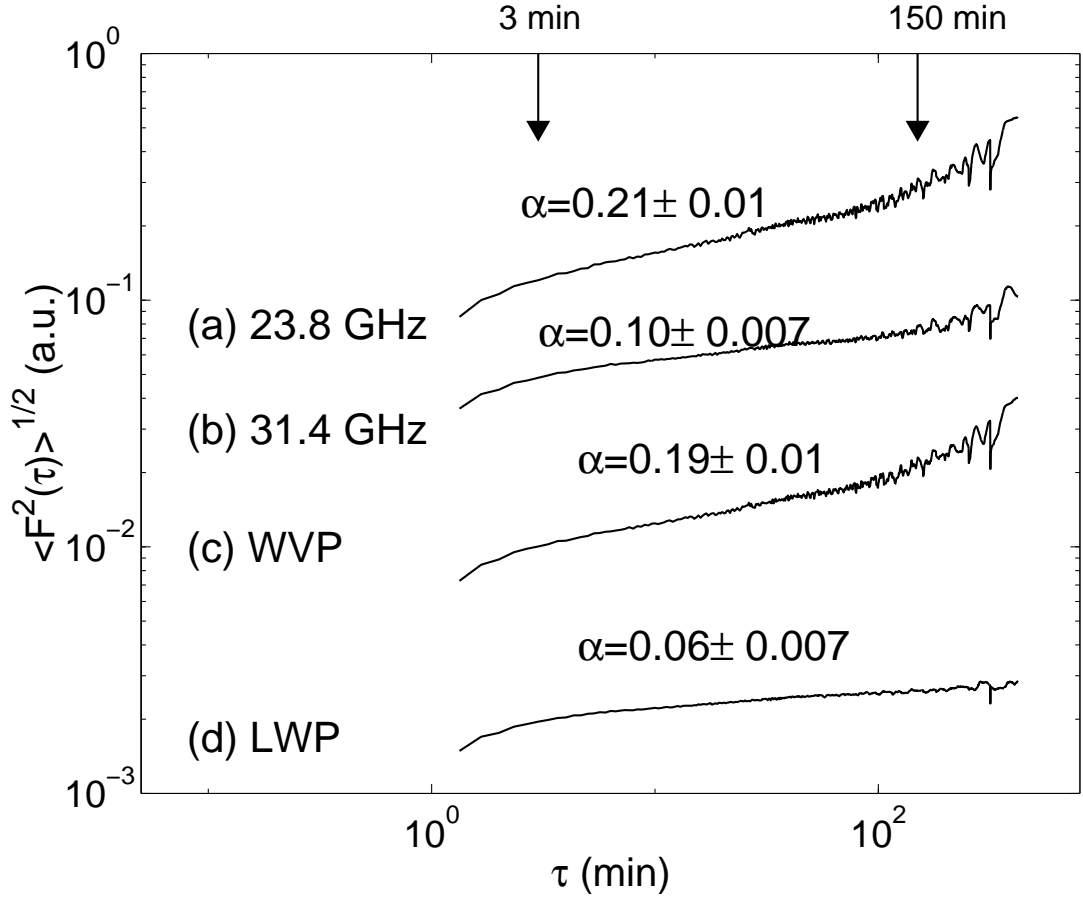}
\end{center}
\caption{The detrended fluctuation analysis (DFA) function $<F^2(\tau)>^{1/2}$ for
the data in Fig. 4. The DFA-function in (b) illustrates that 
the 31.4 GHz brightness temperature channel is dominated by
instrumental noise due to the very low levels of absolute humidity.
The total precipitable water is of order of 0.9~cm.}
\label{fig5}
\end{figure}

\begin{figure}[ht]
\begin{center}
\leavevmode
\epsfysize=12cm
\epsffile{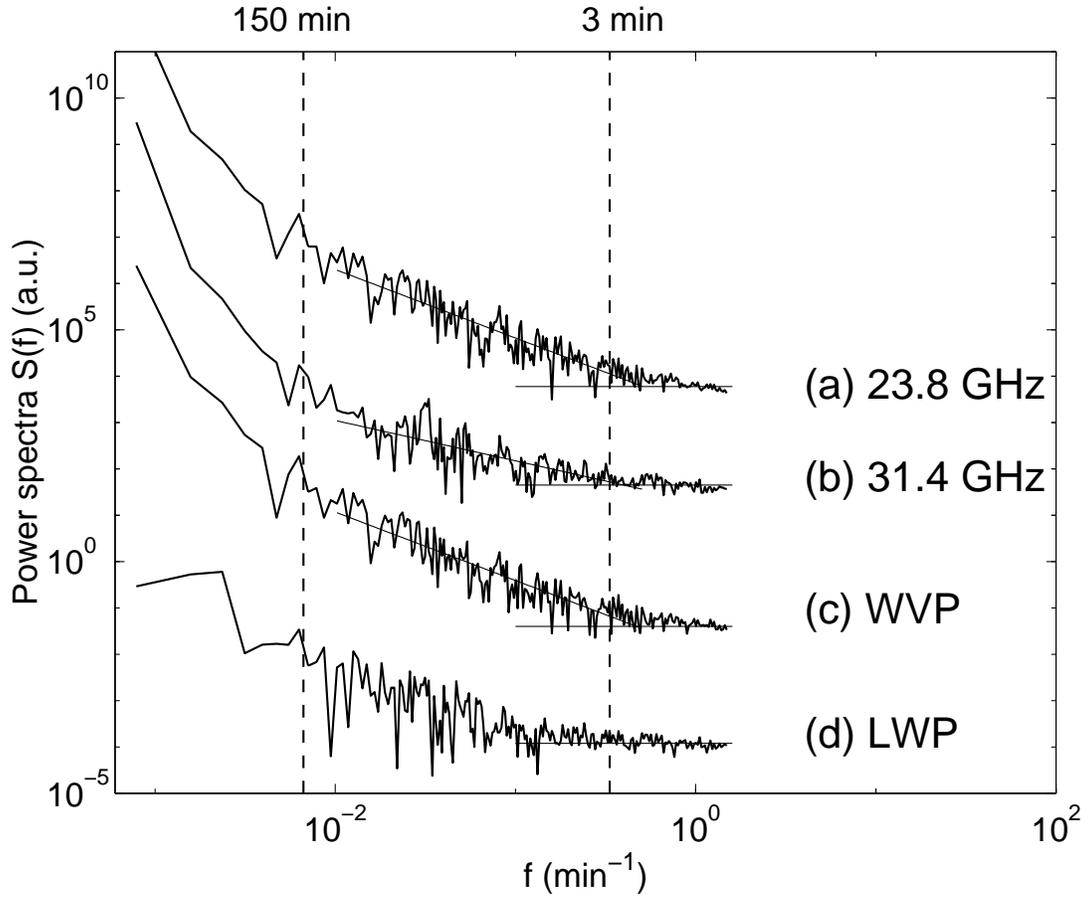}
\end{center}
\caption{Power spectra $S(f)$ for the data in Fig. 4.
The (b) 31.4 GHz channel and (d) LWP spectra suggest that the brightness temperature
records and LWP retrieval at high frequencies for a dry cloudless  atmosphere 
are dominated by the microwave radiometer instrumental noise.}
\label{fig6}
\end{figure}

\begin{figure}[ht]
\begin{center}
\leavevmode
\epsfysize=6cm
\epsffile{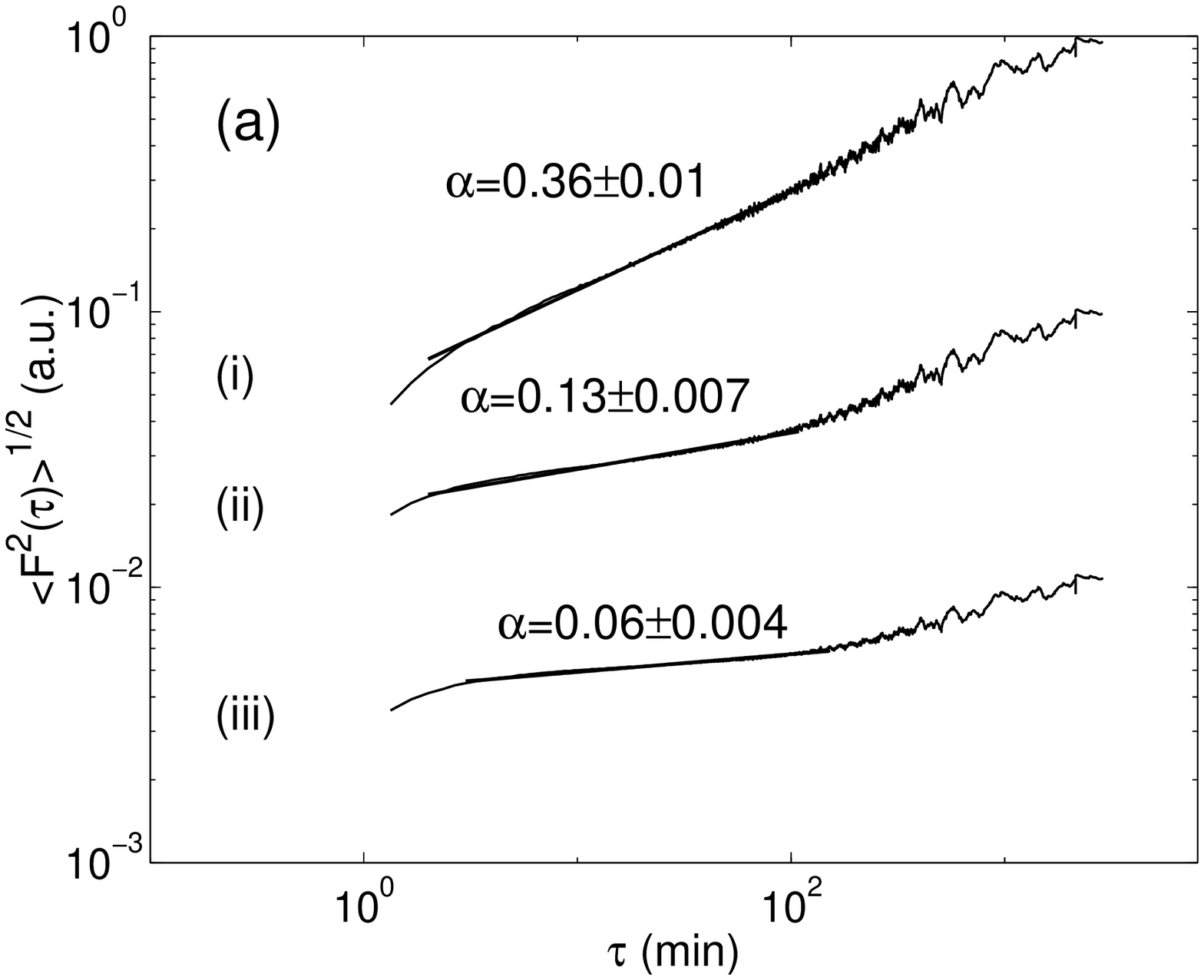}
\vfill
\leavevmode
\epsfysize=6cm
\epsffile{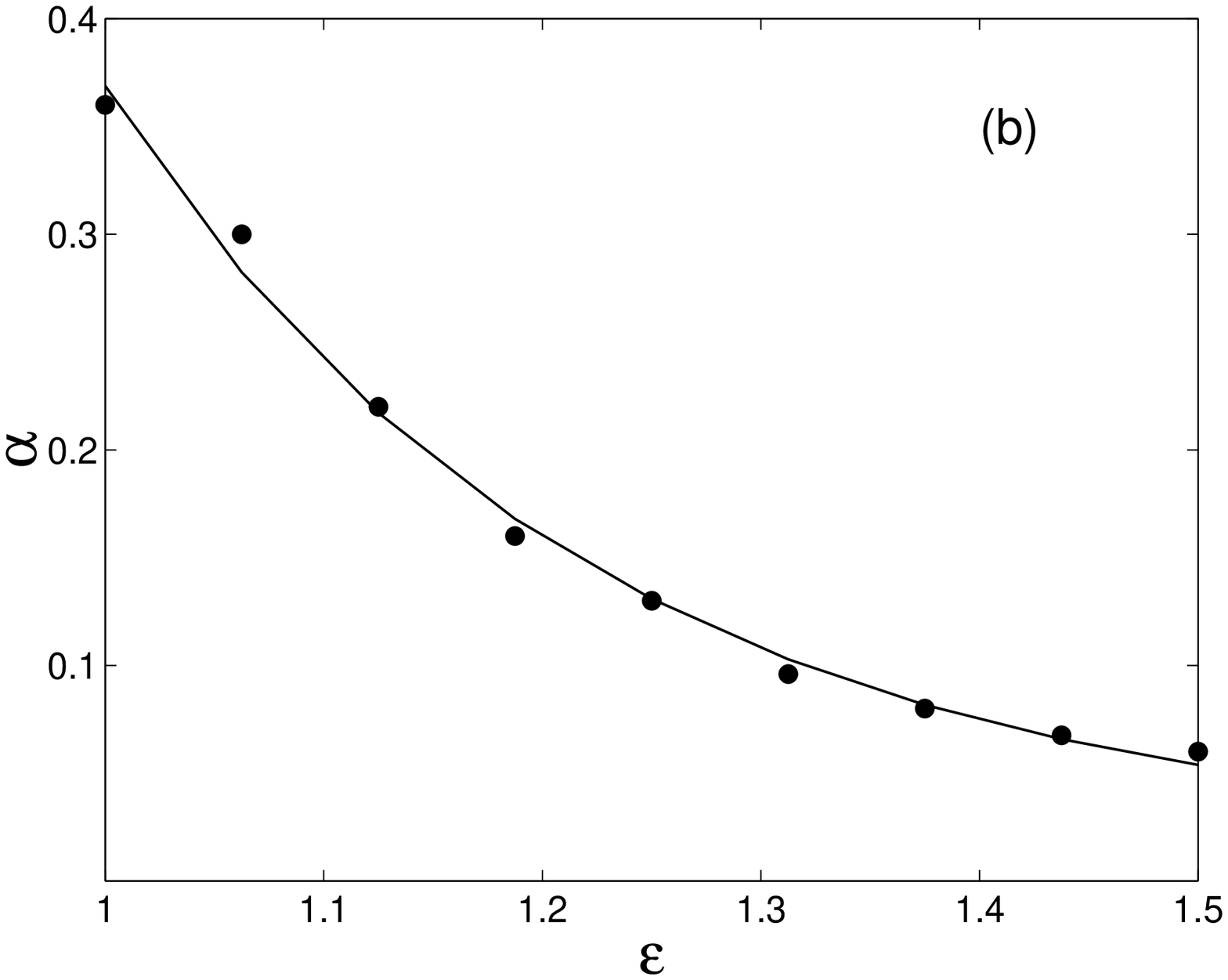}
\vfill
\leavevmode
\epsfysize=6cm
\epsffile{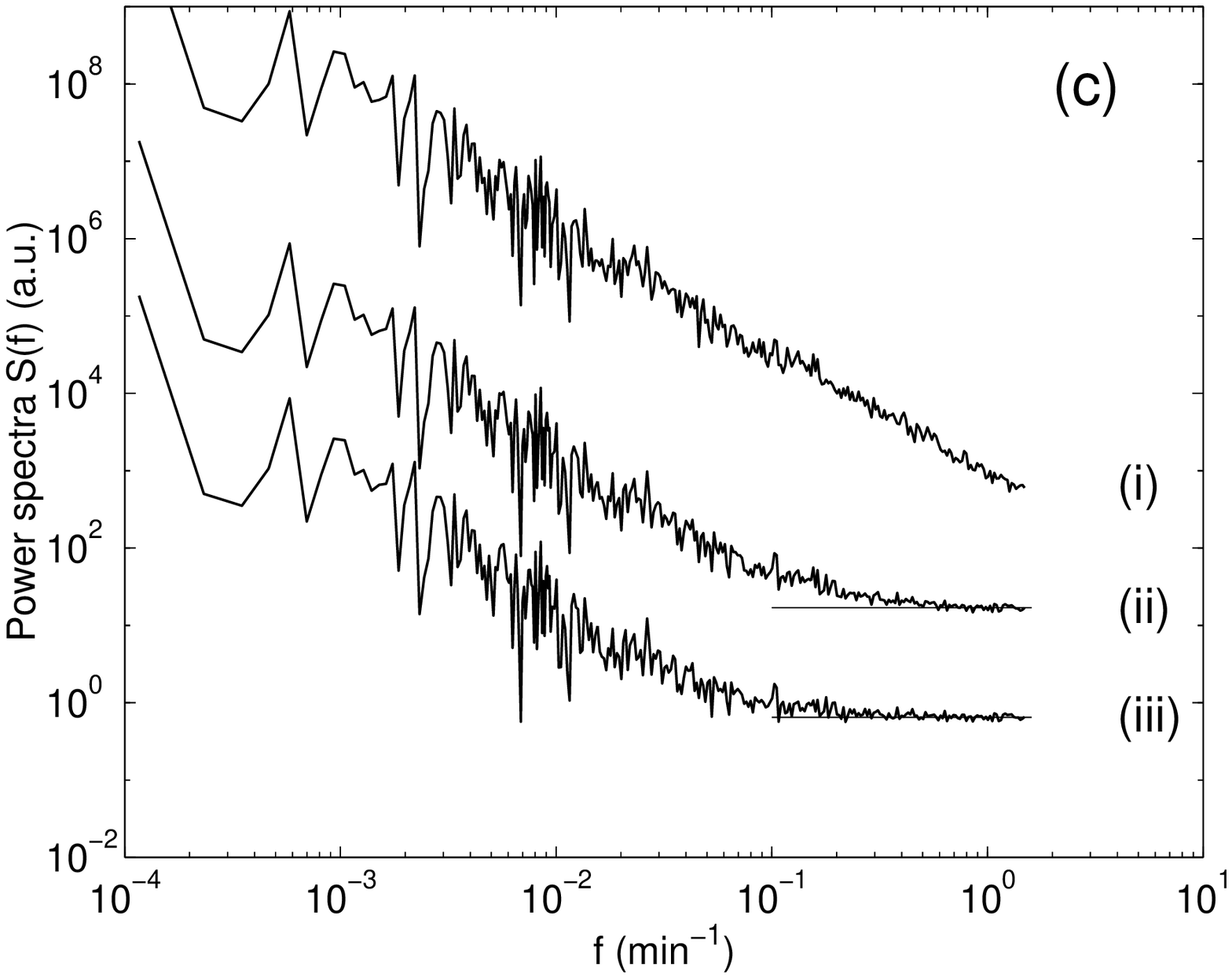}
\end{center}
\caption{(a) DFA-function $<F^2(\tau)>^{1/2}$ for the LWP data illustrated in 
Fig. 1d is labeled (i).
Adding  Gaussian white noise with amplitude of 1/4 and 1/2 of the LWP signal
amplitude to the LWP signal results in the DFA-fluctuation function $<F^2(\tau)>^{1/2}$
labeled (ii) and (iii), respectively.  
(b) The $\alpha$-exponent for the LWP signal (in Fig. 1d) 
with added Gaussian white noise as a function of the noise-to-signal ratio 
$\epsilon$.The beginning of the abscissa corresponds to the LWP signal 
without noise ($\epsilon=1$). Exponential decay is found to be the best 
fit, showing that adding noise to a noisy signal produces an incredibly 
small effect.
(c) Power spectra for the three cases shown in (a), in which the effects of the noise
are clearly seen at high frequencies.}
\label{fig7}
\end{figure}

\begin{figure}[ht]
\begin{center}
\leavevmode
\epsfysize=6.5cm
\epsffile{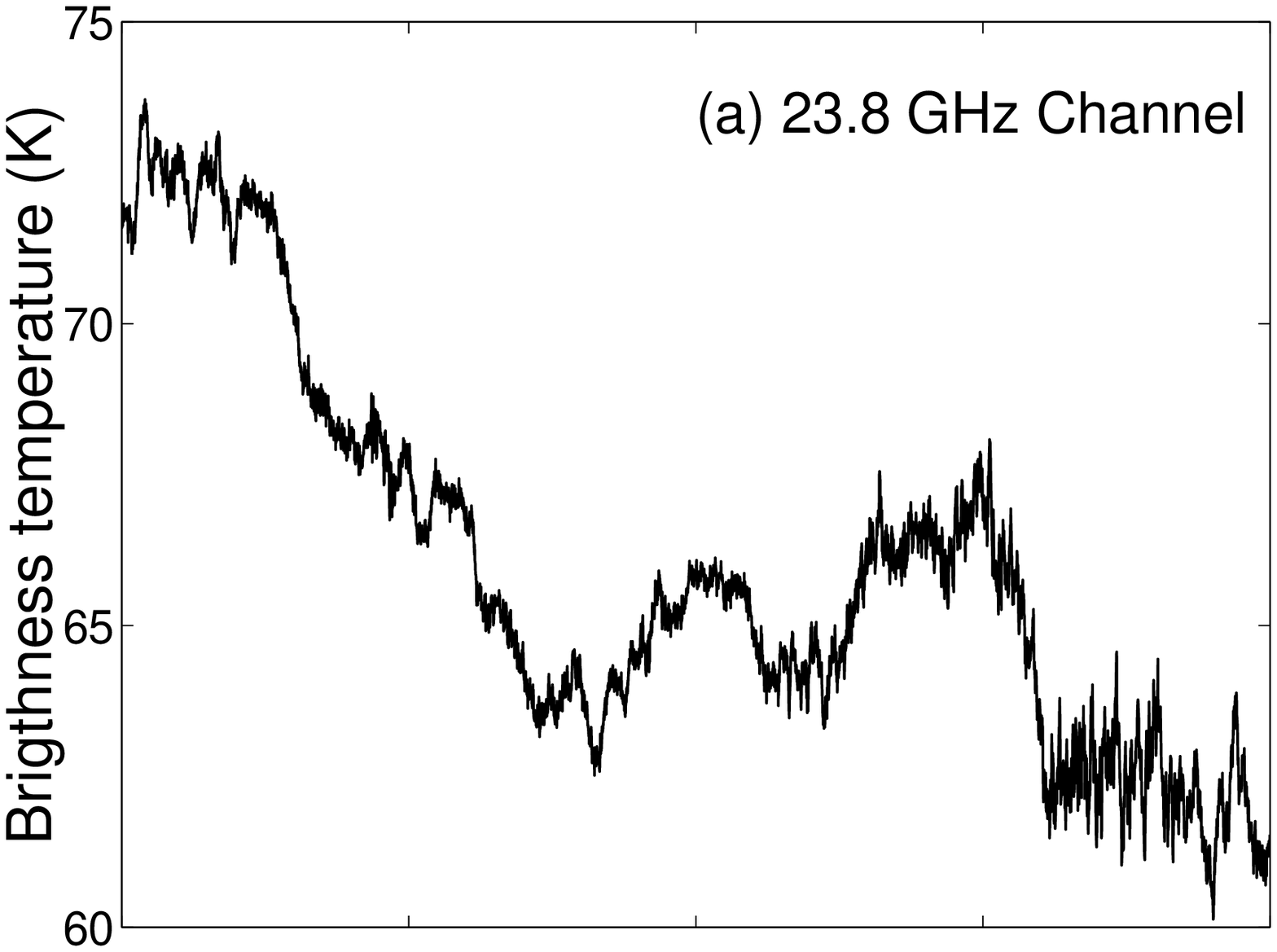}
\hfill
\leavevmode
\epsfysize=6.5cm
\epsffile{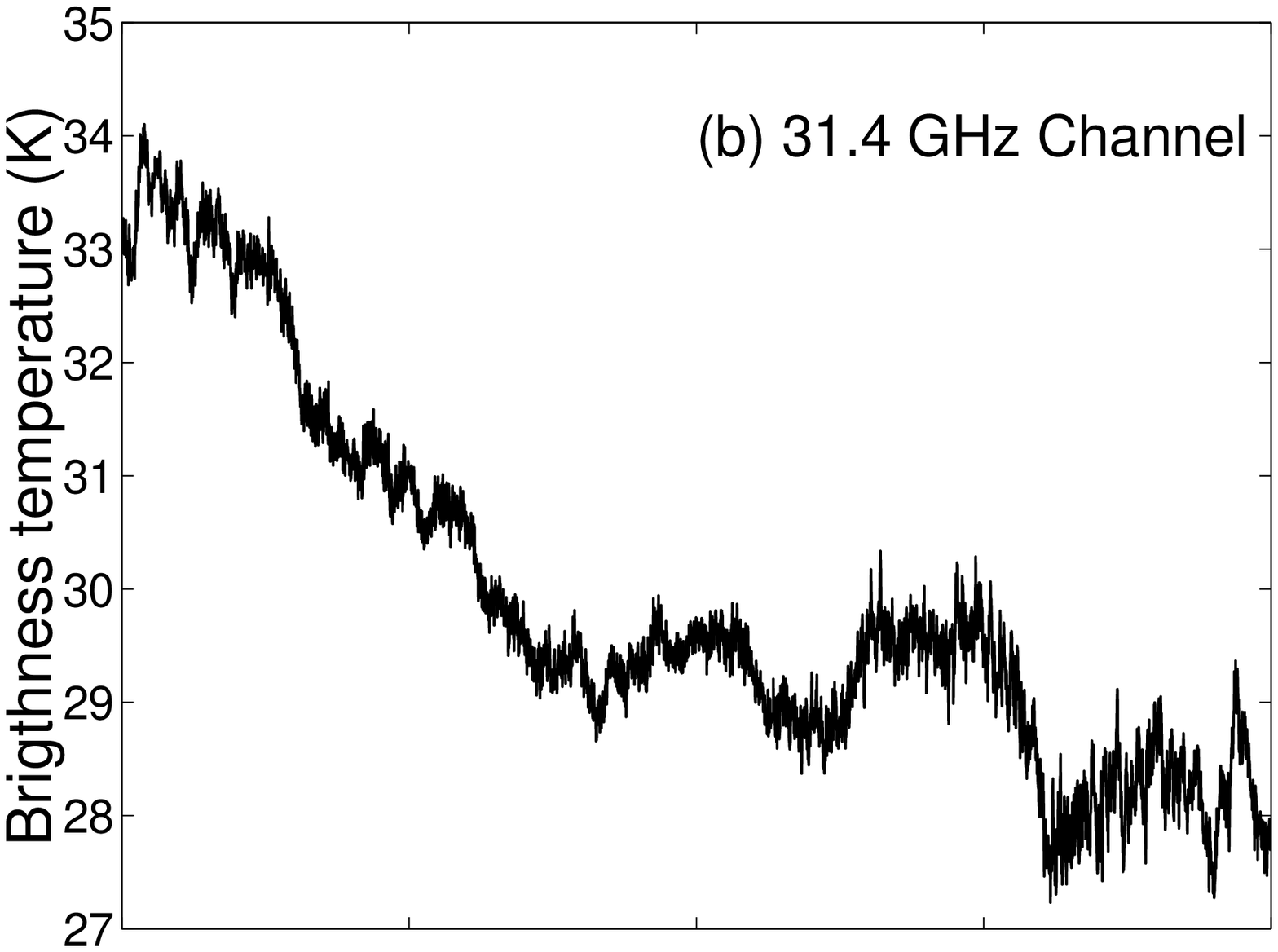}
\vfill
\leavevmode
\epsfysize=7cm
\epsffile{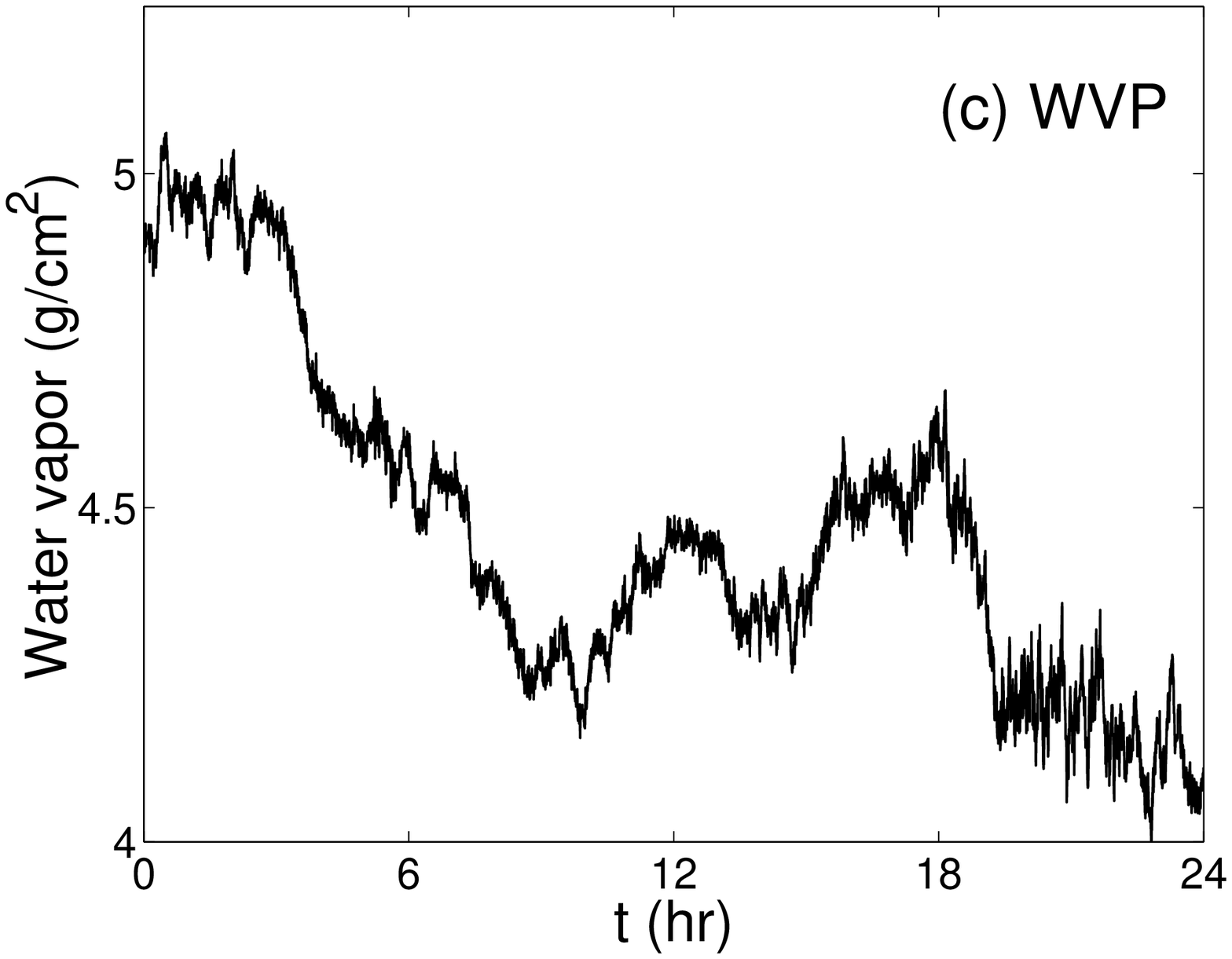}
\hfill
\leavevmode
\epsfysize=7.3cm
\epsffile{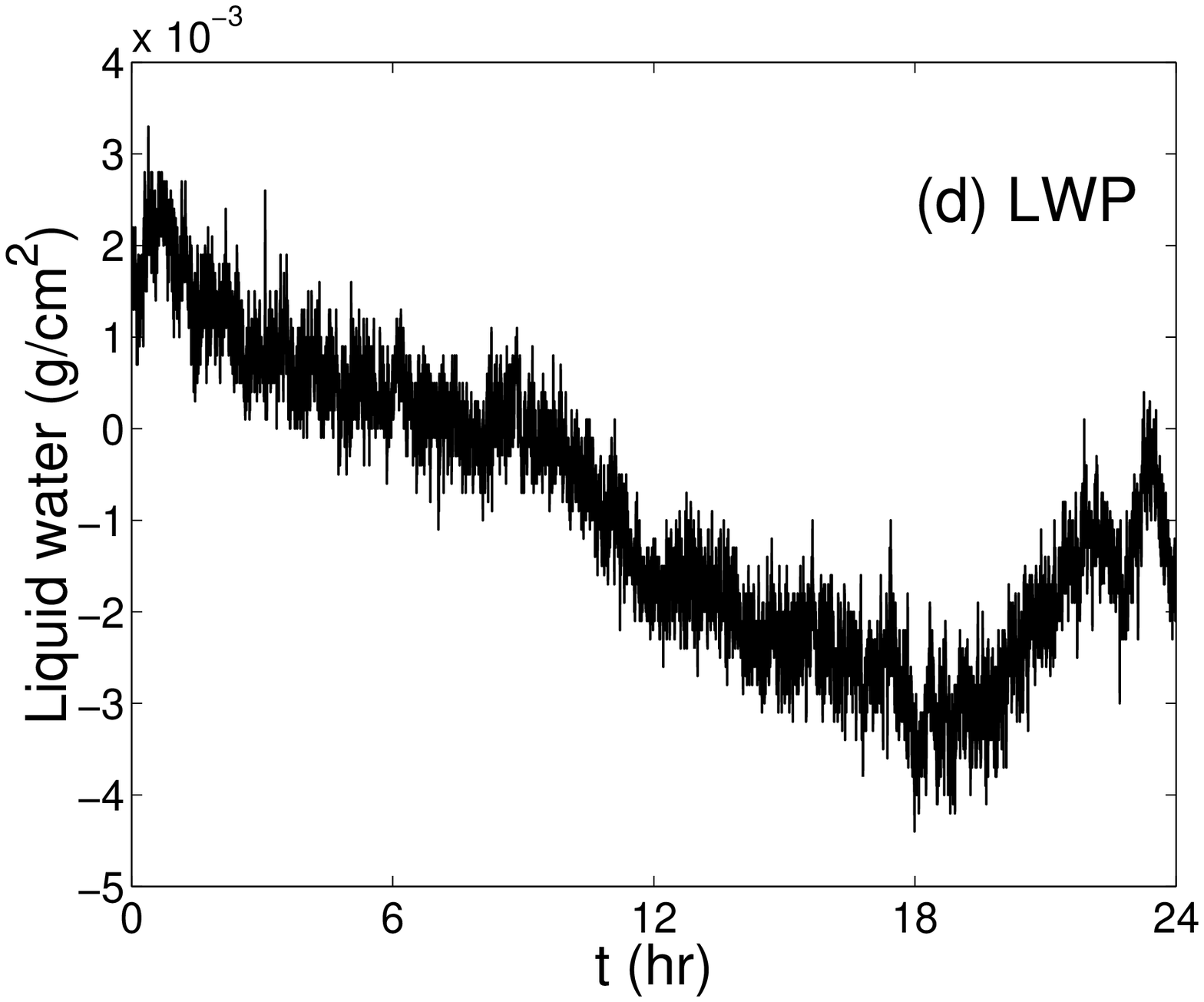}
\end{center}
\caption{Time dependence of the microwave radiometer (a) 23.8 GHz
brightness temperature ${T_B}_1$, (b) 31.4 GHz
brightness temperature ${T_B}_2$, (c) retrieved water vapor path (WVP)
and (d) retrieved liquid water path (LWP) for a moist cloudless  case. 
These measurements were obtained
at the ARM Southern Great Plains site  on September 18, 1997.
Each time series contains $N=4296$ data points with a time resolution 
of 20~s.}
\label{fig8}
\end{figure}

\begin{figure}[ht]
\begin{center}
\leavevmode
\epsfysize=12cm
\epsffile{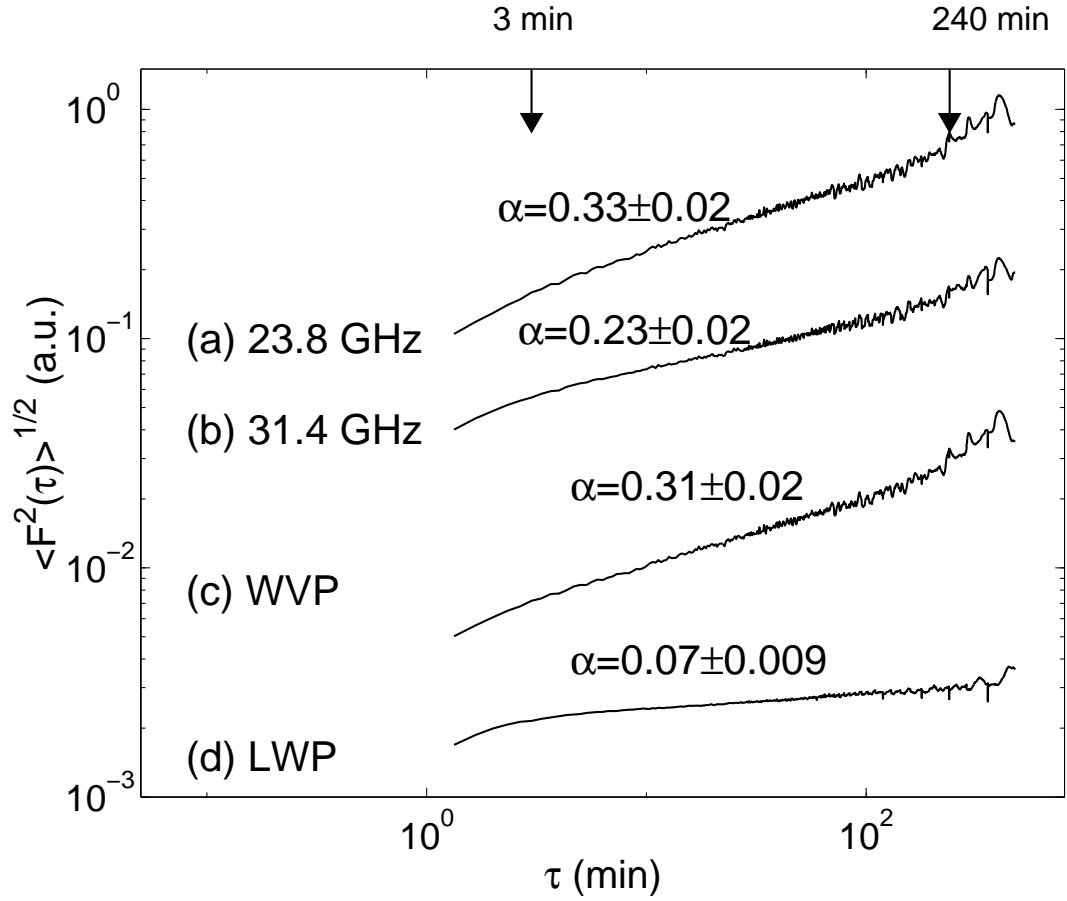}
\end{center}
\caption{The detrended fluctuation analysis (DFA) function $<F^2(\tau)>^{1/2}$ for
the data in Fig. 8. The DFA-function in (d) illustrates that the retrieved LWP
is a  noise-like sequence. This is
an expected result for a cloudless, moist atmosphere.
The total precipitable water is of order of 4-5~cm.}
\label{fig9}
\end{figure}

\begin{figure}[ht]
\begin{center}
\leavevmode
\epsfysize=12cm
\epsffile{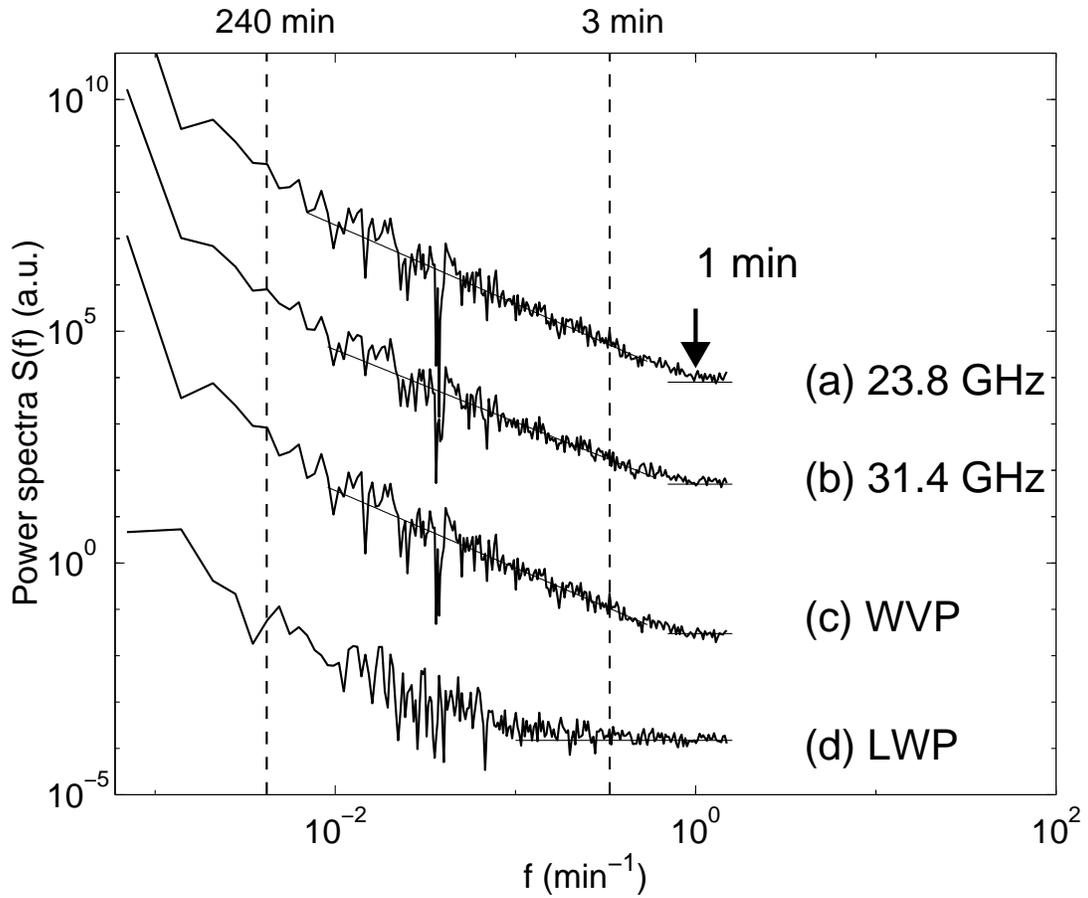}
\end{center}
\caption{Power spectra $S(f)$ for the case of
cloudless, moist atmosphere (data in Fig. 8).
Both 23.8 GHz and 31.4 GHz channels spectra, as well as the 
water path retrieval show noise-like scaling below
1~min.}
\label{fig10}
\end{figure}

\begin{figure}[ht]
\begin{center}
\leavevmode
\epsfysize=8.2cm
\epsffile{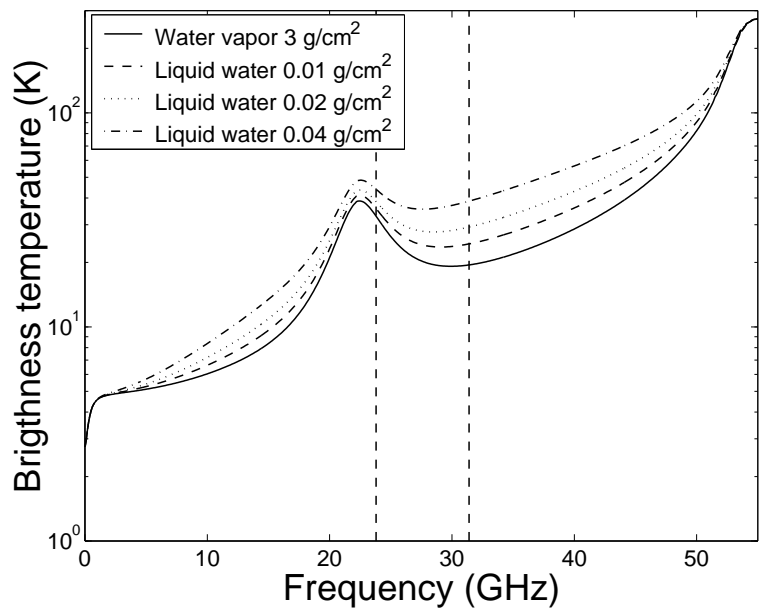}
\end{center}
\caption{Microwave radiometer emission spectra computed with the code of
Schroeder and Westwater, 1991. The solid curve corresponds to the
spectrum of water vapor while the other curves represent the spectra
for different amounts of liquid water in the atmosphere.}
\label{fig11}
\end{figure}

\end{document}